
%
%

\def\unmezzo{{\textstyle 1\over 2}}
\def\dueterzi{{\textstyle 2\over 3}}
\def\O{{\cal O}}

\def\>t{>_{\scriptscriptstyle{\rm T}}}
\def\kint{\int{\d^3k\over(2\pi)^3}}
\def\pint{\int{\d^3p\over(2\pi)^3}}

\def\d{{\rm d}}

\def\r{{\bf r}}
\def\k{{\bf k}}
\def\p{{\bf p}}
\def\q{{\bf q}}
\def\zr{{\bf z}}

\def\v{{\bf v}}
\def\ss{{\bf s}}
\def\l{{\bf l}}
\def\t{{\hat t}}
\def\Ckol{C_{Kol}}
\def\flux{\bar\epsilon}
\def\zq{{\zeta_q}}
\def\b{b_{kpq}}
\def\bun{b^{\scriptscriptstyle (1)}_{kpq}}
\def\bdu{b^{\scriptscriptstyle (2)}_{kpq}}
\def\btr{b^{\scriptscriptstyle (3)}_{kpq}}

\def\z0q{{\zeta^{\scriptscriptstyle{0}}_q}}
\def\smalR{{\scriptscriptstyle R}}
\def\smalze{{\scriptscriptstyle (0)}}
\def\smalun{{\scriptscriptstyle (1)}}
\def\smaldu{{\scriptscriptstyle (2)}}
\def\smaltr{{\scriptscriptstyle (3)}}
\def\smalno{{\scriptscriptstyle (9)}}
\def\smalMAX{{\scriptscriptstyle MAX}}
\def\shell{{\tt S}}
\def\ball{{\tt B}}
\font\brm=cmr10 at 24truept
\font\bfm=cmbx10 at 15truept
\centerline{\brm A closure model for intermittency in three}
\vskip 5Pt
\centerline{\brm dimensional incompressible turbulence}
\vskip 20pt
\centerline{Piero Olla}
\centerline{The James Franck Institute and Ryerson Laboratories,}
\centerline{The University of Chicago, Chicago IL 60637}
\vskip 25pt
\centerline{\bfm Abstract }
\vskip 5pt
\noindent A simplified Lagrangean closure for the Navier-Stokes equation is
used to study
the production of intermittency in the inertial range of three dimensional
turbulence. This is
done using localized wavepackets following the fluid rather than a
standard Fourier basis. In this formulation, the equation for the energy
transfer acquires a noise term coming from the fluctuations in the energy
content of the different wavepackets. Assuming smallness of the intermittency
correction to scaling allows the adoption of a quasi-gaussian approximation for
the velocity field, provided a cutoff on small scales is imposed and a finite
region of space is considered. In this approximations, the amplitude of the
local energy transfer fluctuations, can be
calculated self consistently in the model. Definite predictions on anomalous
scaling are obtained in terms of the modified structure functions:
$<<E(l,a)>_\smalR^q>$,
where $<E(l,a,\r,t)>_\smalR$ is the part of the turbulent energy coming from
Fourier components in a band $(a-1)k$ around $k\sim l^{-1}$, spatially
averaged over a volume of size $R\sim{l\over a-1}$ around $\r$.

\vskip 15pt
\noindent PACS numbers 47.27. -i
\vskip 1cm
\centerline{Submitted to Physics of Fluids, 11-15-1994}
\vfill\eject
\centerline{\bfm I. Introduction}
\vskip 5pt
The statistics of large Reynolds numbers 3D (three dimensional) turbulence is
characterized by scaling behaviors of the structure functions:
$S(l,q)=<v_l^q>$,
$\v_l(\r,t)$ $=\v(\r+\l,t)-\v(\r,t)$, which, in first approximation, appear to
follow the Kolmogorov relation: $S(l,q)=\propto l^\z0q$, $\z0q\simeq q/3$
$[1]$.
However, both experiments $[2]$ and numerical simulations $[3]$ show the
presence
of corrections to Kolmogorov scaling, which become more and more pronounced for
higher order moments. These corrections go in a direction of
increasing nongaussianity of $v_l$ as $l\to 0$: $S(l,q)\propto l^\zq$, with
$\delta\zeta_q=\zq-\z0q$, satisfying the convexity condition:
${\d^2\delta\zeta_q\over\d q^2}<0$.
Of course the relations above hold only for scales corresponding to the
inertial
range (and a part of the viscous range $[4]$ if only ratios of moments are
considered), so that for finite Reynolds numbers: $Re<\infty$, the generalized
kurtosis $K(2q)=[S(l,q)/S(l,2)^q]_{\scriptscriptstyle l\to 0}$ is finite.
However, a situation with scaling corrections in the form just described,
persisting for $Re\to \infty$, would imply that $K(q>2)\to\infty$ in this
limit,
corresponding to infinitely intermittent small scale velocity fluctuations.

The traditional approach to the issue of intermittency dates back to
the remarks by Landau $[5]$ on the presence of spatial fluctuations in the
energy transfer between scales and its effect on the turbulent dynamics.
{}From the Refined Similarity hypothesis of Kolmogorov $[6]$, down the line to
the Beta Model $[7$-$8]$, the main ingredient is the assumption of a turbulent
dynamics acting locally in scale, so that the energy density at a given point
in space is essentially the product of coefficients, each with an independent
fluctuating component, and each describing the transfer of energy between
successive, contiguous scales. In this picture the source of intermittency
lies in the inertial range and acts locally in scale; therefore, the scaling
corrections is predicted to be independent of $Re$. This mechanism can be
interpreted as a rule for the construction of a multifractal $[9$-$10]$ in
which the measure is given by: $d\mu_l(\r,t)=l<(\partial_x v_x)^2>_l(\r,t)$,
where $<\;>_l(\r,t)$ indicates a spatial average taken at time $t$ in a an
interval of length $l$ along the $x$-axis, centered around $\r$. In this way,
the multifractal dimensions $D_q$, defined by the relation $<\d\mu_l^q>\sim
l^{(q-1)D_q}$, are given by: $(q-1)(1-D_q)=q\zeta_2-\zeta_{2q}$.

Recently, an alternative explanation for the presence of scaling
corrections has been proposed, namely that what is observed is
a finite size, i.e. a finite Reynolds number, effect $[11$-$12]$. This
point of view is supported both by the smallness of the corrections and the
strongly nonlocal character of finite size effects with respect to scale, as
shown in numerical
simulations of both Navier-Stokes dynamics $[2]$ and reduced wavevector models
$[11]$. In $[12]$ a dynamical explanation for this effect was presented, based
on closure arguments. The basic idea is that the strong intermittency of the
viscous range, due to the interplay between dissipation and nonlinearity in the
dynamics of small vortices $[13]$, may be enough to generate, in the inertial
range, scaling corrections
of the appropriate size for all Reynolds numbers of practical interest. In this
picture, the total intermittency, as parametrized by $K(q)$, would remain
finite, while the scaling corrections would tend to zero in the infinite
Reynolds number limit.

In this situation, it would be interesting to have some quantitative assessment
of the magnitude of the finite size effects, relative to the amount of
intermittency
produced in the inertial range by energy transfer fluctuations. However, while
$[11]$ and $[12]$ can afford some quantitative prediction on the size of the
first effect, all models dealing with the second are heavily phenomenological
and have no connection with real Navier-Stokes dynamics. An exception is the
very recent paper by Yakhot $[14]$, in which quantitative predictions have been
obtained after approximating the turbulent energy dynamics with  that of a
passive scalar.

The purpose of the present paper is to obtain quantitative predictions on the
$Re$-independent part of anomalous scaling (if present at all), following the
route of evaluating the energy transfer fluctuations from statistical closure
of the Navier-Stokes equation. The basic difficulty in using this technique
is the real space nature of
the quantities that have to be calculated: the moments $S(l,q)$ $[15]$, while
the main component of the turbulent dynamics, the energy transfer towards
smaller scales, is better described in a Fourier basis. Although a real space
closure of the Navier-Stokes equation was derived in $[16]$, we found very
difficult to extend the model to study energy fluctuations, especially when
trying to separate contributions at different scales.

Here we prefer to follow the idea of Nakano $[17]$ and, in a different context,
of Eggers and
Grossmann $[18]$, of using a localized wavepacket representation of the
Fourier-Weierstrass type rather than a global Fourier basis or a purely real
space one. In this way, energy
fluctuations on a scale $R$, instead of being buried in the phase relationship
between Fourier modes separated by $\Delta k\sim R^{-1}$, are described by the
space dependence of the wavepackets. Now, if the wavepackets are centered
around
wavevectors lying in shells of radii $k_n=a^n k_0$, the spatial extension of a
wavepacket at $k\sim l^{-1}$ will be $R(l)\sim l/(a-1)$. This means that a new
object is being used in place of $S(l,q)$ to give a measure of intermittency:
the generalized structure function $S(l,a,2q)=<<E(l,a)>_\smalR^q>$, where
$E(l,a,)=E(l,a;\r,t)$ is the total energy of Fourier modes in the shell
$k_n\sim l^{-1}$, and $R=R(l,a)=l/(a-1)$. More in general we shall consider the
situation in which
also the quantity $c_w={l\over(a-1)R}$ is treated as a free parameter,
resulting
in the definition of still another structure function $S(l,a,c_w,q)$
interpolating between the intermittency free limit $S(l,a,0,q)$ and the
original
case: $S(l,a,c_w^{\scriptscriptstyle{\rm MAX}},q)\equiv S(l,a,q)$.

The shell width $(a-1)k$ plays here a crucial role; this can be understood
better by looking at how
energy is transferred between shells as a rule for the construction of a
multifractal. In this picture, a small value for $a-1$ corresponds to
multipliers between one scale $l$ and the next $l/a$ being constant on domains
$R(l)>l$ (in typical examples like the various ''middle third'' Cantor sets
$[10]$, one has $R(l)\simeq l$, i.e. the fluctuation scale and that of the
geometrical structures is the same).
In terms of turbulent dynamics, this has the following interpretation: $R(l)$
is the scale of the eddies which contribute the most to the straining of those
of size $l$; the parameter $l^{-1}R(l)=(a-1)^{-1}$ gives therefore the degree
of nonlocality of the nolinear interaction. In principle there will be
fluctuations in the transfer of energy between eddies of size $l$ also at scale
smaller than $r(l)$ and this effect will result in correlations between phases
of different wavepackets separated by $\Delta k>R^{-1}$. This effect
contributes
to the scaling of $S(l,q)$, but not to that of $S(l,a,q)$, and the last one is
likely to underestimate the actual value of $\delta\zq$.

Now, recent analysis performed by Domaradzki and Rogallo $[19]$ on numerical
simulations and statistical closures of the Quasi Normal Markovian type $[20]$
has shown that there is indeed a separation of scales between straining flow
and strained eddies, so that fluctuations in the transfer, occurring at scales
similar to those of the modes exchanging energy, are not expected to be large,
implying: $S(l,q)\sim S(l,a,q)$ even for $a$ close to 1.
More importantly, it is this separation of scales that allows to consider
meaningful wavepackets, extending over several characteristic wavelengths,
rather than having to deal with the usual eddy breaking picture, that is very
attractive on grounds of simplicity, but does not allow to make any contact
with
the Navier-Stokes dynamics.

There is a second conceptual difficulty in using closures to study
multifractal intermittency. This is the contradiction between an ansatz of
quasi gaussianity and the ''infinitely non gaussian'' character of a
multifractal with no cutoffs, as attested by the equation: $K(q>2)=\infty$.
A quasi gaussian hypothesis becomes meaningful however when
studying limited regions of space and limited ranges of scales, which is
possible if the degree of nonlocality of the nonlinear interactions is not too
high and the intermittency correction $\delta\zeta_q$ is sufficiently small.
In this
way, although the distribution of values of, say $\partial_xv_x$, over
different averaging volumes $V_i$ is infinitely intermittent
(if the total volume $V_{tot}=\cup V_i$ is infinite itself), the moments of
$\partial_xv_x$ in each $V_i$ will be close to gaussian $[$if the ultraviolet
cutoff scale $r$ is not too small: $(rV_i^{-1/3})^{\delta\zq/q}-1\ll 1]$. Of
course,
in order for the statistical sample to be significant, it is necessary that
the range of scales $r<l<V_i^{1/3}$ be large enough to accomodate all relevant
interactions: $(a-1)V_i^{1/3}/r>1$, and that enough wavepackets be present in
the volume $V_i$: i.e. $({k\over (a-1)})^3V_i\gg 1$. With a degree of
nonlocality
$(a-1)^{-1}$ of at most ten (see $[19]$ and results in the next section),
and ${\delta\zeta_q\over q}$ in the range of a few
percents $[2]$, these conditions can be assumed to be satisfied,
and all statistical averages, indicated by
$<\;>$, will be understood here to be carried on in space, over a single large
but finite volume $V_i$.

The closure that is going to be used in the next sections is of the
Quasi-Lagrangean type $[21]$, in which fluid structures at scale $l$ are
studied in a reference frame
moving with a speed given by the average fluid velocity in a ball of
radius $l/\lambda$, $\lambda<1$; the free parameter $\lambda$ is then adjusted
to lead, in a mean field theory, to values of the Kolmogorov constant in
agreement with experiments. This mean field theory is obtained using still a
global Fourier basis. Notice that here $l/\lambda$ is not the wavepacket
size, and in this respect, the present approach differs from that of Nakano
$[17]$. Of course our approximation is completely uncontrolled,
in that sweep effects from scales between $l$ and $l/\lambda$ are still
present,
while part of the strain from scales larger than $l/\lambda$ is lost. However,
it is still an improvement upon using infrared cutoffs in the expression for
the
energy transfer, especially in view of the fact that this last operation would
not preserve the nonlocal character of the nonlinear interaction and would lead
to transfer profiles in disagreement with $[19]$.

The fluctuating dynamics is obtained studying the energy equation for the
wavepackets $[$which, after angular integration, becomes the energy equation
for the shells $k_n<k<k_{n+1}]$. This is obtained by substituting the
statistical averages which would lead to a mean field closure, with partial
spatial averages over wavepackets volumes. This leads to an energy equation in
the same form (in Fourier space) as the original mean field one, plus a noise
term that is essentially:
$<v^\smalze(v^\smalze\nabla)v^\smalze>_\smalR(\r,t)$.
Here $v^\smalze$ is the gaussian random field, which gives the lowest order
approximation for the fluid velocity. The end result is an equation for the
energy content of shells at a certain position in space, which is in a form
very close to the stochastic chains studied by Eggers in $[22$-$23]$.

This paper is organized as follows. The closure technique is going to be
described in section II and equations for the energy balance in laboratory
as well in Lagrangean frame are derived in the mean field approximation.
In section III, this closure is applied to the analysis of the energy transfer
fluctuations. In section IV, an energy balance equation in terms of shells is
derived and its solution is used to obtain the intermittency corrections to
scaling. Section V is devoted to discussion of the results and to concluding
remarks. Most of the calculation details have been confined to the appendices.

\vskip 20pt
\centerline{\bfm II. Mean energy transfer in Lagrangean reference frame}
\vskip 10pt
\centerline{\bf A. Closure outline}
\vskip 5pt
In the inertial range of fully developed 3D turbulence, the dynamics obeys the
Euler equation:
$$
\partial_t\v(r,t)+(\v\cdot\nabla)\v(\r,t)+\nabla P(\r,t)=0,\eqno(1)
$$
while the presence of dissipation is accounted for by the energy flux towards
small scales; the pressure $P$ is calculated through the incompressibility
condition $\nabla\cdot\v=0$. In a Lagrangean reference frame, Eqn. (1) can be
rewritten in the equivalent form:
$$
D_t\v(\zr_t+\r,t)=((\v(\zr_t,t)-\v(\zr_t+\r,t))\cdot\nabla)\v(\zr_t+\r,t)
-\nabla P(\zr_t+\r,t)=0,\eqno(2)
$$
where $D_t=\partial_t+\v(\zr_t,t)\cdot\nabla$ is the material derivative
along the trajectory $\zr_t=\zr_t(\r_0,t_0)$ of a Lagrangean tracer that at
time
$t_0$ was at $\r_0$.

The first assumption of the model is that the velocity field can be taken to
lowest order to obey gaussian statistics, with correlation given, in a
Lagrangean frame (with corrections to the exponential decay in $t$ for small
$t$), by the expression:
$$
C_k^{\alpha\gamma}(t)\simeq
\int\d^3r\,e^{i\k\cdot\r} C^{\alpha\gamma}(\r,t;0,0)
$$
$$
=2\pi^2\Ckol\flux^{2/3}P^{\alpha\gamma}(\k)k^{-11/3}\exp(-\eta_k t).
\eqno(3)
$$
$$
C_{\alpha\gamma}(\r,t;\r_1,t_1)\equiv<v^\smalze_\alpha(\zr_t+\r,t)
v^\smalze_\gamma(\zr_{t_1}+\r_1,t_1)>,
$$
where $t>0$, $\Ckol$ is the Kolmogorov constant, $\flux$ the mean energy
dissipation,
$P^{\alpha\gamma}(\k)=\delta^{\alpha\gamma}-{k^\alpha k^\gamma\over k^2}$ the
transverse projector and $\eta_k=\rho\flux^{1/3}k^{2/3}$ the eddy turnover
frequency at scale $k^{-1}$, with $\rho$ a dimensionless constant. The
statistical
average $<\;>$ taken in Eqn. (3) can be understood as an average over the
initial position $\r_0$, which is taken in a large but finite volume, as
discussed in the introduction.

The first nongaussian correction $\v^\smalun$ is obtained from (2); this is
integrated, keeping second order terms, in the Direct Interaction Approximation
(DIA) $[24$-$26]$, rather than following $[16]$. The result contains then a
green function: $G(t,\r|t_1,\r_1)$, which gives the effect of a source, which
at time $t_1$ was in $\zr_{t_1}+\r_1$, on a point $\zr_t+\r$ at time $t$:
$$
\eqalignno{v^\smalun_\alpha(\r_0+&\r,0)=\int_{-\infty}^0\d t\int\d^3 s\,
G_\alpha^\rho(0,\r|t,\r+\ss)\cr
&\times[v^\smalze_\sigma(\zr_t+\r+\ss,t)-
v^\smalze_\sigma(\zr_t,t)]\partial^\sigma v_\rho^\smalze(\zr_t+\r+\ss,t).
&(4)\cr}
$$
The time decay of the (retarded) green function introduced above is assumed to
be the same as that of the correlation function $[26]$ and the following
approximation is adopted:
$$
G(t,\r|t_1,\r_1)\simeq G(t,\r-\r_1|t_1,0)\equiv G(\r-\r_1,t-t_1),\eqno(5)
$$
which means that the initial and final position in $G$ are shifted together
until the first lies on the Lagrangean trajectory  $\zr_t(\r_0,t_0)$.
Eqns. (5) allows a great simplification in taking Fourier transforms, in that
all correlations become diagonal: $<v_{\bf p}v_{\bf q}>\propto\delta({\bf p}
-{\bf q})$. In particular, it becomes possible to write time correlations in
the form: $C^{\alpha\gamma}_k(t)=G_{\beta,k}^\alpha(t)C_k^{\beta\gamma}(0)$,
where:
$$
G_k^{\alpha\gamma}(t)=\int\d^3r e^{i\k\cdot\r}G^{\alpha\gamma}(\r,t)=
P^{\alpha\gamma}(\k)\exp(-\eta_k|t|).\eqno(6)
$$
The meaning of this approximation is that the divergence of Lagrangean
trajectories is disregarded. In particular, the statistical average over
the initial position $\r_0$ coincides identically with spatial average at
the given time. We have also the result that, in this approximation, the
advanced green function coincides  with the ''transpose'', with respect to the
space slots, of the retarded one; in the following sense: for $t_1>t_2$,
$$
<v^\smalze_\alpha(\zr_{t_1}+\r_1,t_1)v^\smalze_\gamma(\zr_{t_2}+\r_2,t_2)>=
\int\d^3 s G^\sigma_\gamma(t_1,\r_2+\ss|t_2,\r_2)
C_{\alpha\sigma}(\r_2+\ss-\r_1,0)
$$
$$
=\int\d^3 s G^\sigma_\alpha(t_1,\r_1|t_2,\r_1+\ss)
C_{\sigma\gamma}(\r_1+\ss-\r_2,0).\eqno(7)
$$
Further discussion of these points is contained in Appendix A.

The final assumption is necessary in order to separate sweeping from
straining scales, and is that, inside averages, one has to carry on the
following substitution:
$$
<...[\v(\zr+\r)-\v(\zr)]\cdot\nabla...>_k\to<...[\v(\zr+\r)-\hat w(\lambda,k)
\v(\zr)]\cdot\nabla...>_k,\eqno(8)
$$
where $\hat w(\lambda,k)$ is a smoothing operator acting on $\v(\zr)$ by
filtering out Fourier modes above $\lambda k$. The hypothesis here is that,
although all velocity components are integrated along a single Lagrangean
trajectory $\zr_t(r_0,t_0)$, when these components are large scale ones, small
scale details of $\zr_t$ will not contribute in averages.
\vskip 10pt

\centerline{\bf B. Mean field analysis}
\vskip 5pt
In order to study the fluctuation dynamics in the energy transfer $T$, we will
have
to work in a Lagrangean frame. However, in order to fix the free parameter
$\lambda$ introduced above, it is necessary to match theoretical predictions
with experimental data of the Kolmogorov constant that are taken in a fixed
laboratory frame. For this reason, the first step in the analysis is the
derivation of closure equations for eulerian correlations.

The calculations to obtain the energy equation in the laboratory frame are
standard $[26]$:
by multiplying Eqn. (1) by $\v(\r_0,0)$ and taking the average, the pressure
term drops off because of incompressibility, and one is left with:
$$
\partial_t C^{\alpha\gamma}(\r,t)|_{t=0}=
-<v^\alpha(0,0)v^\beta(\r,0)\partial_\beta v^\gamma(\r,0)>,\eqno(9)
$$
which of course is equal to zero at steady state. Actually, this is the
equation for
the Eulerian 2-time correlation at zero time separation; the energy equation is
obtained from (9) by multiplying its RHS (right hand side) by 2.
Expansion to first order in $v^\smalun$ and use of Eqns. (3-4) and (7),
leads to the same expression for the energy equation as
in DIA (both Eulerian and Lagrangean, $[24$-$26]$), and in ''Quasi
Normal-Markovian'' closures $[20]$:
$$
\left.{\partial C_k(t)\over\partial t}\right|_{t=0}=\Big({\pi\over k}\Big)^2
T(k)= {1\over 4\pi^2}\int_\Delta
\d p\d q\,kpq\,\theta_{kpq}\b C_q(C_p-C_k),\eqno(10)
$$
where: $C_k^{\alpha\gamma}(t)=P^{\alpha\gamma}(\k)C_k(t)$; $C_k\equiv C_k(0)$;
$\Delta$ is the domain defined by the triangle inequalities: $p>0$ and
$|k-p|<q<k+p$;
$\theta_{kpq}=(\eta_k+\eta_p+\eta_q)^{-1}$ is the relaxation time; $\b=
(p/k)(xy+z^3)$ is the geometric factor, in which $x,y$ and $z$ are cosines
of the angles opposite respectively to $k,p$ and $q$ in a triangle with sides
$kpq$. The terms associated with integrating along $\zr$ are uniform in space,
they do not couple with the others and are shown in Appendix A not to
contribute to the final result: in this model, the choice of integrating back
in time along a Lagrangean path is felt only in the eddy turnover time
$\theta_{kpq}$.
\vskip 4pt
The green function $G$ is a Lagrangean object and is obtained by multiplying
this time Eqn. (2) by $\v(\r_0,0)$ and then taking averages. The resulting
triplet term is in the form:
$<v^\alpha(\r_0,0)[v^\beta(\zr_t+\r,t)-v^\beta(\zr_t,t)]\partial_\beta
v^\gamma(\zr_t+\r,t)>$; now, $v^\beta(\zr_t,t)$, that is the term coming from
the shift to a Lagrangean frame, contributes to the final expression and is
responsible for the cancellation of the sweep terms. Substituting Eqns. (3-4)
and (7) in the new triplet, and expanding again to first order in $v^\smalun$,
we obtain at steady state the following equation:
$$
{DC_k(t)\over Dt}={1\over 4\pi^2}\int_0^t\d\tau\int_\Delta
\d p\d q[\b G_p(t-\tau)C_k(\tau)C_q(t-\tau)
$$
$$
+w_p(\lambda,k)\bun G_p(t-\tau)C_k(t-\tau)C_q(\tau)
-w_q(\lambda,k)\bdu G_k(t-\tau)C_p(\tau)C_q(t-\tau)]\eqno(11)
$$
where: $G_k^{\alpha\gamma}=P^{\alpha\gamma}G_k$,
$\bun=(p/2k)(xy(1-2z^2)-y^2z)$ and $\bdu=(p/2k)(xy+z(1+z^2-y^2))$ are
new geometric terms, and $w_q(\lambda,k)$ gives the effect in $k$-space of the
cutoff operator $\hat w(\lambda,k)$
The term in Eqn. (11), which cancels the divergence of the integral
for $q\to 0$, is the one in $\bdu$. Integrating from $t=0$ to
$t=\infty$ and using Eqns. (3) and (6), we get the result:
$$
{\rho^2\over \Ckol}=\unmezzo\int_\Delta\d p\d q\,pq^{-8/3}\,\Big[
{\b\over p^{2/3}+q^{2/3}}+{w_p\bun\over(1+p^{2/3})q^{2/3}}-
{w_q\bdu\over(1+q^{2/3})p^3}\Big],\eqno(12)
$$
which gives a first equation connecting $\Ckol$ and $\rho$ with $\lambda$.
A second equation connecting $\Ckol$ and $\rho$, given an energy balance
equation in the form of (10), was derived by Kraichnan $[27]$:
$$
\rho/\Ckol^2\simeq 0.19.\eqno(13)
$$
The constant $\Ckol$ considered in this section is a well defined quantity,
provided the average volume $V_i$ and the range of scales $k$ are not too
large;
in this sense we have locally: $C_k(V_i)\propto\Ckol\flux_i^{2/3}k^{-11/3}$,
even though for $V_i\to\infty$ anomalous corrections become important, and
$\Ckol$ ceases to have a clear meaning.
Imposing that the Kolmogorov constant matches the experimentally observed value
$\Ckol\approx 1.5$ (we are assuming here that that the finite size effects and
intermittency corrections which affect the experimental $\Ckol$ are indeed
small), and using a gaussian profile for the cutoff $w$:
$w_p(\lambda,k)=\exp(-(p/\lambda k)^2)$, Eqns. (12) and (13) set:
$\lambda=0.9$.
In the following we will fix therefore:
$$
w_p(\lambda,k)\to w_p(k)=\exp(-1.23(p/k)^2).\eqno(14)
$$
The same steps leading from Eqns. (1) to (10) can be repeated starting from
Eqn. (2). The result is the following equation for velocity correlations in a
Lagrangean frame:
$$
\left.{DC_k(t)\over Dt}\right|_{t=0}=\Big({\pi\over k}\Big)^2T(\lambda',k)=
$$
$$
{1\over 4\pi}\int_\Delta\d p\d q\,kpq\,\theta_{kpq}[\b+w_q(\lambda',k)\btr]
C_q(C_p-C_k),\eqno(15)
$$
where: $\btr=(p/k)(1+xy+z(z^2-{x^2+y^2\over 2}))$ $[28]$. Again, the
corresponding energy equation is obtained by multiplying the RHS of (15) by 2
and understanding the $t$ on the LHS (left hand side) not as a
time separation in a 2-time correlation but as the $t$-dependence in a non
stationary 1-time correlation. Notice that the parameter $\lambda'$ entering
Eqn. (15) does not
have to coincide with $\lambda=0.9$, which is fixed and separates
between sweeping and straining scales. This allows to calculate the Kolmogorov
constant and the parameter $\rho$ in different reference frames. The result
is shown in Fig. 1. and suggests that, although differences are
expected between quantities measured in laboratory and Lagrangean frame (due to
correlation among reference frame velocity and quantity to be averaged), their
orders of magnitude should be the same. This is particularly important because
the intermittency estimates that are going to be derived in the next sections
will have to be calculated in a Lagrangean, not a laboratory frame.
\vskip 4pt
The structure of the energy transfer is studied, following Domaradzki and
Rogallo, $[19]$ by decomposing $T(\lambda,k)$ in its contribution from
different
scales:
$$
T(\lambda,k)=\int\d p\,T(\lambda,k,p).\eqno(16)
$$
It appears that the present closure is able to maintain the features of large
scale straining observed in $[19]$ also in a Lagrangean frame, as it should be
expected. It is clear instead, from Fig. 2., that simpler closures
based on a Navier-Stokes nonlinearity in $k$-space, amputated of the large
scale convection contributions:
$$
(k_\alpha P_{\beta\gamma}(\k)+k_\gamma P_{\beta\alpha}(\k))
v^\alpha_{\bf p}v^\gamma_{\bf q}\to
(k_\alpha(1-w_q)P_{\beta\gamma}(\k)+k_\gamma(1-w_p)P_{\beta\alpha}(\k)))
v^\alpha_{\bf p}v^\gamma_{\bf q},\eqno(17)
$$
would lead to transfer profiles almost without any exchange of energy between
nearby scales.


\vskip 20pt
\centerline{\bfm III. Energy transfer fluctuations}
\vskip 5pt
Although the analysis in section II disregarded fluctuations, all averages were
implicitly dependent on the position (through $\flux$), at the scale of the
volumes $V_i$. In this section, we calculate the same averages over balls of
radius $R$ and the variations of the result from ball to ball
are used to study the fluctuations of the energy transfer. If the ratio
$RV_i^{-1/3}$ is not too large, these fluctuations are going to be small and
the analysis can be carried on in perturbation theory.

In principle, it would be nice to carry on the calculations in the laboratory
frame,
where experimental data are available. However, the statistics of the velocity
field is ill defined there, due to the effect of sweep. The balls are then
imagined to move rigidly along Lagrangean trajectories passing through their
centers at time $t_0$, while, thanks to the simplifying assumptions of Eqns.
(6) and (7), the average over the initial position is substituted by a spatial
average inside the balls. As discussed in the introduction, we associate to
each scale $l$, a certain radius $R(l)\sim l/(a-1)$; this automatically
induces a basis of wavepackets of width $\Delta k\sim R^{-1}$. In particular,
we
obtain a partition of Fourier space in shells $\shell_n$ of radii $k_n=a^nk_0$,
each
containing $\sim(k_n/\Delta k)^2$ wavepackets, associated with the different
orientations of the wavevector $\k$ $[17]$. Clearly, a further degeneracy in
the wavepackets is produced by their different location in real space.

For $R/l$ large, the main interactions occur among overlapping balls and are
associated with local transfer of energy towards small scales. We consider
therefore a sequence of nested balls and study the transfer of energy among
them. Notice that a derivation of deterministic equations for a shell model
of the type considered here, would require some dynamical analog of the
statistical assumptions of Eqns. (5-7); at this point it is more natural to
follow the route of statistical closure to the end.

Let us indicate with $<\;>_m$ the spatial average over $\ball_m$: the ball of
radius $R_m$ associated with the $m$-th shell; once we have fixed the origin of
the axis of the moving frame on $\zr_t$, we can write $<\;>_m$ in terms of a
kernel $W(r,m)$:
$$
<\Psi>_m=\int\d^3 r\,W(r,m)\Psi(\r)=\kint e^{i\k\cdot\r}W_k(m)\Psi_\k,
$$
In this way, the energy density in $\ball_m$ reads:
$$
E(m)=\int\d k\,E_k(m);\qquad E_k(m)=2\pi k^2\pint W_p(m)\v_{{{\bf p}\over
2}+\k}
\cdot\v_{{{\bf p}\over 2}-\k},\eqno(18)
$$
To fix our ideas we shall consider gaussian wavepackets:
$$
W_k(m)=\exp(-(k/\Delta k_m)^2),\qquad\Delta k_m=c_w(a-1)k_m,\eqno(19)
$$
with $c_w$ relating the wavepacket thickness $\Delta k_m$ and the shell spacing
$k_{m+1}-k_m=(a-1)k_m$. Notice that $c_w$ can be treated as a free parameter,
in that, for a fixed choice of shell (i.e. for a given value of $a$), we can
still consider arbitrarily thin wavepackets, or in the limit, even global
Fourier modes. Of course, it is the thickest wavepacket for a given $a$ (that
is the maximum value of $c_w$), which will be able to catch most of the
fluctuation dynamics.

If $k_nR_m$ is large, $W_p(m)\propto R_m^{-3}\delta(\p)$ and we can
approximate:
$$
E_k(m)\simeq (1+\phi_n(m))E^\smalze_k\equiv(1+\phi_n(m))\Ckol\flux^{2/3}
k^{-5/3},\eqno(20)
$$
with $E_k^\smalze=C_k k^2/2\pi^2$ the spectral energy density in $V_i$,
$n=n(\k)$ the shell of the wavevector $\k$, and $\phi_n(m)=\phi_n(m,t)$
fluctuating and small.

The energy density $E(m)$ can be expressed also as a sum of contributions from
different shells:
$$
E(m)=\sum_n E_n(m);\qquad E_n(m)=\int_n\d kE_k(m).
\eqno(21)
$$
The term $E_n(m)\sim<E(k_n^{-1},a,\zr_t,t)>_m$ can be seen as the istantaneous
total energy of wavepackets in $\shell_n$, lying in the volume $\ball_m$ at
$\zr_t$; however, if $\ball_m$ becomes smaller than the wavepackets, i.e. (for
$c_w$ maximal) $n>m$, the only average taking place will be over wavevector
orientations and will be independent of $m$; hence we get:
$\phi_n(m<n)=\phi_n(n)$.

With these definitions, the energy equation for the shells reads:
$$
D_tE_n(m)=T_n(m)+f_n(m).\eqno(22)
$$
In the equation above, the energy transfer into $\shell_n$, averaged over
$\ball_m$, has two components. The first:
$$
T_n(m)=\int_n\d k T'(k),\eqno(23)
$$
is a relaxation term giving the response of the system to fluctuations in the
energy content of the various shells. For small $\phi$, the integrand $T'(k)$
coincides with the transfer term $T(\lambda,k)$ of Eqn. (15), with the
substitution: $C_k\to ( 1+\phi_n(m))C_k$.
For $c_w\to c_w^\smalMAX$, the $\d k$ integral in Eqn. (23) receives
discreteness
corrections from the shell and the wavepacket thickness being comparable;
notice however that the $\int_\Delta\d p\d q$ integral contained in $T'(k)$
$[$see Eqns. (10) and (15)$]$ remains unaffected, when $a-1$ is small.

The second term contains the fluctuations of the transfer, around the average.
To lowest order in the expansion around gaussian statistics it has itself two
components. The first comes from the pressure: $<\v\cdot\nabla P>_m$
and results in a surface integral over the boundary of $\ball_m$. The
second: $<\v^\smalze\cdot (\v^\smalze\cdot\nabla)\v^\smalze>_m$ did not
contribute in the mean field analysis of the previous section, because of the
gaussianity of $\v^\smalze$; it plays a role here however, by acting as a
source of fluctuations in Eqn. (22).
Now, we are considering a situation in which $kR$ is large; hence, the pressure
contribution can be neglected and we are left with:
$$
f_n(m)=-<\hat h(n)[\v^\smalze(\zr_t,t)\cdot
((\v^\smalze(\zr_t+\r,t)-
\v^\smalze(\zr_t,t))\cdot\nabla)\v^\smalze(\zr_t+\r,t)]>_m,\eqno(24)
$$
where $\hat h(n)$ is a band pass filter for the modes in $\shell_n$:
$\hat h(n)f(\r)\equiv\int\d^3r'h(n,r')f(\r+\r')=
\kint h_k(n)f_\k e^{i\k\cdot\r}$, with $h_k(n)=H(k-k_n)
-H(k-k_{n+1})$; $H(x)$ is the Heaviside step function.

Treating $f_n(m)$ as an external (nongaussian) noise, Eqn. (22) becomes a
stochastic differential equation for the $\phi_n(m)$'s. Following Eggers
$[22$-$23]$, we express $f_n(m)$ as the difference between the fluctuations in
the energy flux across $k_n$ and $k_{n+1}$:
$$
f_n(m)=g_n(m)-g_{n+1}(m),\eqno(25)
$$
where $g_n$ and $g_{n+1}$ are associated one with each of the step functions
entering the definition of $h_k(n)$.
\vskip 10pt

\centerline{\bf A. Fluctuation source}
\vskip 5pt
The statistics of $g_n(m)$ can be calculate explicitly from Eqns. (3) (24) and
(25). Here we shall content ourselves with the analysis of the 2-point
correlations. Already, this calculation requires the evaluation of some fifteen
contractions of the product:
$$
v^\alpha(\zr_1)[v^\beta(\zr_1+\r_1)-v^\beta(\zr_1)]\partial_\beta
v^\alpha(\zr_1+\r_1)v^\gamma(\zr_2)[v^\sigma(\zr_2+\r_2)-v^\sigma(\zr_2)]
\partial_\sigma v^\gamma(\zr_2 +\r_2);\eqno(26)
$$
these contractions are listed in table B1. in Appendix B. Of these only six
contribute; one example is contraction (B1.9):
$$
<g_{n_1}(m_1,t_1)g_{n_2}(m_2,t_2)>^\smalno=
$$
$$
\int\d^3 r_1\d^3 r_2H(r_1,n_1)H(r_2,n_2)\int\d^3 z_1\d^3
z_2W(z_1,m_1)W(z_2,m_2)
C^{\alpha\gamma}(\zr_1-\zr_2,t)
$$
$$
\times\partial_\sigma[C^{\beta\gamma}(\zr_1+\r_1
-\zr_2-\r_2,t)-\hat wC^{\beta\gamma}(\zr_1-\zr_2-\r_2,t)]
$$
$$
\times\partial_\beta[C^{\alpha\sigma}(\zr_1+\r_1-\zr_2-\r_2,t)
-\hat wC^{\alpha\sigma}(\zr_1+\r_1-\zr_2,t)];\eqno(27)
$$
notice the cutoff operators $\hat w$ signaling a term coming from working in a
Lagrangean reference frame. In terms of Fourier components, we get:
$$
<g_{n_1}(m_1,t_1)g_{n_2}(m_2,t_2)>^\smalno=
$$
$$
\int{\d^3 k\over(2\pi)^3}{\d^3 p\over(2\pi)^3}{\d^3 q\over(2\pi)^3}
C_kC_pC_q\exp(-(\eta_k+\eta_p+\eta_q)|t|)
$$
$$
\times W_{\p+\q-\k}(m_1)W_{\p+\q-\k}(m_2)\,pkyz(y+xz)
$$
$$
\times[(2-w_p(k))H(k-k_{n_1})-w_p(k)H(q-k_{n_1})]
$$
$$
\times[(2-w_q(k))H(k-k_{n_2})-w_q(k)H(p-k_{n_2})].\eqno(28)
$$
(The wavevectors entering the two last lines of Eqn. (28) can be tracked back
to Eqn. (26) and (B1.9): the $p$ and $k$ in $w_p(k)$ come from
$v^\beta_p$ and $v^\alpha_k$, while the $k$ and  $q$ in $H(k-k_{n_1})$ and
$H(q-k_{n_1})$ come from $v_k^\alpha$ and $v_q^\alpha$).
It has already been mentioned that, for large $kR$, the averaging kernel $W$
is proportional in $k$-space to a Dirac delta. Also the product $W(m_1)W(m_2)$
is a Dirac delta:
$$
W_\k(m_1)W_\k(m_2)\simeq(c_w(a-1))^3\Big({\pi\,k_{m_1}^2k_{m_2}^2\over
k_{m_1}^2+
k_{m_2}^2}\Big)^{3/2}\,\delta(\k),\eqno(29)
$$
where use has been made of Eqn. (19). Eqn. (29) allows simplification of Eqn.
(28) by means of the
bipolar integral formula $[26]$: $\int\d^3p\d^3q\delta(\p+\q-\k)={2\pi pq\over
k}
\int_\Delta \d p\d q$. Repeating the calculations leading to (28) with all the
other contractions we obtain the result:
$$
<g_{n_1}(m,t_1)g_{n_2}(m,t_2)>={\pi^{1\over 2}\flux^2\over 2^{9\over 2}}
[\Ckol\,c_w\,(a-1)k_m]^3
$$
$$
\times\int_0^\infty\d k\int_\Delta\d p\d q\,(kpq)^{-8/3}
\exp(-(\eta_k+\eta_p+\eta_q)|t|)
$$
$$
\times[H_1B_1(kpq)+H_2B_2(kpq)+H_3B_3(kpq)],\eqno(30)
$$
The terms $B_i$ $i=1,2,3$ are geometrical factors similar to the $b_{kpq}$
terms entering the expression for the transfer function.
The factors $H_i=H_i(n_{1,2};k,p,q)$ are expressed in terms of step functions
and restrict the integrals to the appropriate domains for the calculation of
Lagrangean frame energy fluxes; they have the same origin as the two last lines
of Eqn. (28). The exact form of both the $B_i$ and $H_i$
functions is given in Eqns. (B4-7) of Appendix B.

For $t=0$, Eqn. (30) can be reduced to double integrals using the standard
change of variables $[26]$: $k=k_n/u$, $p=k_nv/u$, $q=k_nw/u$ and exploiting
the similarity of the integrand with respect to $u$. Similarly, the following
expression for the noise correlation time:
$$
\bar\eta_n(m)^{-1}=<f_n(m)^2>^{-1}\int_0^\infty\d t<f_n(m,t)f_n(m,0)>,
$$
can be reduced, after explicit calculation of the time integral, to a double
integral. Numerical evaluation gives then the results:
$$
<g_n(m)^2>\simeq 0.15\,(\Ckol\,c_w\,(a-1))^3a^{l(n,m)}\flux^2;\eqno(31)
$$
and:
$$
\bar\eta_n(m)=\bar\rho\flux^{1/3}k_n^{2/3}\qquad\bar\rho\simeq 0.6,\eqno(32)
$$
where $l(n,m)=\min(0,m-n)$ for $c_w$ maximal. The decay of the correlation with
respect to scale is shown in Fig. 3. Notice that $\bar\rho\simeq
2\rho_{\scriptscriptstyle\lambda=0.9}$ (see Fig. 1.), which is what one would
get identifying
naively: $<g(0)g(t)>\propto<v(0)v(t)>^2$. Similarly, the decay of $<g_ng_{n'}>$
with respect to $(n-n')\ln a$ is approximately the same as that of the
auto-correlation for $|T(k,p)|$: $A(q)=\int\d p|T(k,p)T(k+q,p+q)|$, which is
due to the fact that the energy flux at scale $k$ is roughly equal to
$\int\d p|T(k,p)|$.

\vskip 10pt
\centerline{\bf B. Relaxation term}
\vskip 5pt
The shell energy equation (22) is in the form of a nonlinear stochastic
differential equation. Since $\phi$ is small, we can linearize $T_n(m)$
(including the frequencies $\eta_k\sim k^{3\over 2}E_k^{1\over 2}$ entering
the term $\theta_{kpq}$),
with a result that is similar to the stochastic model of Eggers
$[22]$. One difference is the presence here of energy transfer between non
adjacent shells. To analyze this issue, we write the transfer $T_n(m)$ as
a sum of contributions in which the wavevectors $k$ $p$ and $q$ are
respectively in $\shell_n$, $\shell_r$ and $\shell_s$:
$$
T_n(m)={\Ckol^2\flux\over\rho}\sum_{r,s}T_{nrs}(m);\qquad
T_{nrs}=c_{nrs}\phi_r+c_{nsr}\phi_s-d_{nrs}\phi_n,\eqno(33)
$$
where, from Eqns. (15) and (20):
$$
c_{nrs}=\int_n\d k\int_{\Delta_{rs}}\d p\d q\,\tilde\theta_{kpq}
\Big[\tilde a_{kpq}k^2(pq)^{-5/3}-\Big[\dueterzi p^{2/3}\tilde\theta_{kpq}
\Big(\tilde a_{kpq}k^2(pq)^{-5/3}
$$
$$
-\tilde b_{kpq} p^2(kq)^{-5/3}-\tilde b_{kqp}
q^2(kp)^{-5/3}\Big)+\tilde b_{kpq}q^2(kp)^{-5/3}\Big]\Big],\eqno(34)
$$
and:
$$
d_{nrs}=\int_n\d k\int_{\Delta_{rs}}\d p\d q\,\tilde\theta_{kpq}
\Big[\dueterzi k^{2/3}\tilde\theta_{kpq}
\Big(\tilde a_{kpq}k^2(pq)^{-5/3}
$$
$$
-\tilde b_{kpq} p^2(kq)^{-5/3}-\tilde b_{kqp}
q^2(kp)^{-5/3}\Big)+\tilde b_{kpq}p^2(kq)^{-5/3}+b_{kqp}q^2(kp)^{-5/3}\Big].
\eqno(35)
$$
The quantities appearing on the RHS of Eqns. (33-34) are defined as follows:
$\Delta_{rs}$ is the restriction of $\Delta$ to $p\in\shell_r$;
$q\in\shell_s$; $\tilde b_{kpq}=\b+w_q(k)\,\btr$; $\tilde a_{kpq}=\tilde
b_{kpq}
+\tilde b_{kqp}$ and $\tilde\theta_{kpq}=(k^{2/3}+p^{2/3}+q^{2/3})^{-1}$.
The RHS of Eqns. (33-34) can be reduced to double integrals using the same
method of Eqn. (30), and are evaluated numerically.
We can rearrange the sum in Eqn. (33) in the following form:
$$
\sum_{r,s}T_{nrs}(m)=\sum_r a_r(m)\phi_r(m)
=\sum_{r=1}^\infty A_r(m)\Delta^r\phi_n(m),\eqno(36)
$$
where $\Delta$ is the finite difference operator acting as follows:
$$
\Delta^{2r+1}\phi_n={1\over 2}(\Delta^{2r}\phi_{n+1}-\Delta^{2r}\phi_{n-1});
$$
$$
\Delta^{2r}\phi_n=\Delta^{2(r-1)}\phi_{n+1}+\Delta^{2(r-1)}\phi_{n-1}-
2\Delta^{2(r-1)}\phi_n,\eqno(37)
$$
so that the coefficients $A_r(m)$ are expressed in terms of the $a_r(m)$'s
through the relation:
$$
A_{2r+1}=a_{n+2r-1}-a_{n-2r+1};\qquad A_{2r}=
{a_{n+2r-1}+a_{n-2r+1}\over 2}\eqno(38)
$$
The fact that only differences between $\phi_n$ enter the expression above is
due to the fact that we are expanding around a Kolmogorov spectrum, for which
$T_n=0$. For the same reason, the coefficients $c_{nrs}$ and $d_{nrs}$ are
invariant under the transformation: $\{n,r,s\}\to\{n+j,r+j,s+j\}$, which
explains the fact that the finite difference coefficients $A_r$ do not depend
on the shell index $n$.

In Fig. 4 these coefficients are plotted in terms of the shell constant $a$.
Notice that finite difference of order higher than $2$ appear to be negligible
for most choices of $a$.  We obtain therefore the basic result that the shell
dynamics obeys, for most
values of $a$, a (discrete) heat equation forced by a random noise, with an
advection term proportional to $\Delta$.

The physical picture corresponding to this result is not new $[29]$.
Fluctuations over finite volumes $\ball_m$ moving with the flow,
in the energy transfer to eddies at scale $k_n$, generate
fluctuations in the energy content of these volumes and scales.
The eddies are stretched by the turbulent flow, so that their energy
(together with its fluctuation $\phi_n$) is transferred towards smaller scales;
the term responsible for this effect is the advection $A_1\Delta\phi_n$. At the
same time, the randomness of the turbulent flow causes some eddies to be
stretched more and some less, resulting in the end in a diffusion of energy
over different scales.
\vskip 20pt

\centerline{\bfm IV. Anomalous scaling estimates}
\vskip 5pt
The fluctuations in the energy of wavepackets at different space locations is
what is responsible for intermittency in this model.
Instead of studying the scaling behaviors of the structure functions $S(l,q)=
<v^q_l>$, we focus on the modified structure functions, defined through:
$$
S(k_n^{-1},a,c_w,2q)= <<E_n>_n^q>=E_n^\smalze<(1+\phi_n(n))^q>.
\eqno(39)
$$
Identifying naively the space separation $l$ in $S(l,q)$ with the wavelength
$k_n^{-1}$ in Eqn. (39), the two definitions of structure function coincide,
for an appropriate choice of $a$, and for $c_w=c_w^\smalMAX$.
Given a large enough averaging volume $V_i$, the intermittency correction is
small and we can linearize in $n$; in this way, using also: $k_n=k_0a^n$, we
have:
$$
\zeta_{2q}-q\zeta_2\simeq-{1\over\ln a}{\d\over \d n}<(1+\phi_n(n))^q>.
\eqno(40)
$$
If we confine ourselves to the lowest order moments, large fluctuations
of $\phi$ do not contribute too much, so that we can expand:
$(1+\phi_n(n))^q\simeq 1+{q(q-1)\over 2}<\phi_n(n)^2>$. Imposing the Kolmogorov
relation for the third order structure function: $S(l,a,3)\propto l$, fixes
the value for $\delta\zeta_2$, leading to the lognormal statistics result
$[6]$:
$$
\delta\zeta_q\simeq-{q(q-3)\over 4\ln a}{\d<\phi_n(n)^2>\over\d n};
\eqno(41)
$$
we see then that the presence of anomalous scaling is associated with secular
behavior of the fluctuations $\phi_n(n)$.

Unfortunately, the energy balance for the wavepackets, Eqn. (22) is not an
equation for $\phi_n(n)$, but one for $\phi_n(m)$ for fixed $m$. We can rewrite
Eqn. (22) in a more explicit form, by approximating the relaxation term
$T_n(m)$ with the advection-diffusion operator introduced through Eqns.
(35-37),
and by
rewriting the noise term $f_n$ as a difference of energy flux fluctuations at
different scales, as from Eqn. (25). After adequate rescaling, we are left
with the equation:
$$
[\exp(-\gamma n)\partial_\t-D\Delta^2+V\Delta)]\phi_n(m,t)=\hat\Delta
(F_n(m)^{1/2}\xi),\eqno(42)
$$
where $\hat\Delta\xi_n=\xi_{n+1}-\xi_n$ and:
$$
\t=t\,\bar\rho\flux^{1/3}k_0^{2/3};\qquad\gamma={2\over 3}\ln a;\qquad
D={\Ckol\over \rho\bar\rho(1-a^{-2/3})}A_2;
$$
$$
V={\Ckol\over\rho\bar\rho(1-a^{-2/3})}A_1;\qquad
F\simeq 0.83{(c_w(a-1))^3\Ckol\,a^{l(m,n)}\over(\bar\rho(1-a^{-2/3}))^2};
$$
$$
<\xi_{n+{n'\over2}}(m,\t)\xi_{n-{n'\over2}}(m,0)>\simeq{\exp(-e^{\gamma n}|\t|)
\over 1+112.5(\gamma\,n')^2}.\eqno(43)
$$
Here the Kolmogorov and time scale constants $\Ckol$ and $\rho$ are the ones
measured in the moving frame: $\Ckol\simeq 0.8$ and $\rho\simeq 0.3$ (it
appears however that all final results do not change by more than $10\%$ by
exchanging these values with their Eulerian counterparts: $\Ckol\sim 1.5$ and
$\rho\simeq 0.43$). The dependence of the noise correlation $<\xi\xi>$ on
$n-n'$ in the formula above is a fit of the result of numerical integration
shown in Fig. 3.

Clearly, Eqn. (42) does not lead to intermittent behaviors; the factor
$a^{l(m,n)}=a^{(m-n)}$ in the noise amplitude, which goes to zero at small
scales prevents it. Passing from Eqn. (42) to an equation for $\phi_n(n)$
requires the introduction of corrections due to the fact that now,
wavepackets associated with different scales do not overlap exactly in real
space. There are two such contributions:
$$
(-D\Delta^2+V\Delta)\phi_n(m)|_{m=n}
-(-D\Delta^2+V\Delta)\phi_n(n)
$$
$$
=(-D+V/2)(\phi_{n+1}(n+1)-\phi_{n+1}(n))\eqno(44a)
$$
and
$$
\hat\Delta(F^{1/2}(n,m)\xi_n(m))|_{n=m}
-\hat\Delta(F^{1/2}(n,n)\xi_n(n))=
$$
$$
F^{1/2}(n,n)(\xi_{n+1}(n+1)-\xi_{n+1}(n)).\eqno(44b)
$$
The shell equation for $\phi_n\equiv\phi_n(n,t)$ takes then the form (to
simplify notations, the hat on the rescaled time $\t$ will be dropped in the
following):
$$
[\exp(-\gamma n)\partial_t-D\Delta^2+V\Delta)]\phi_n=
F_n^{1/2}(\hat\Delta\xi+\delta\xi),\eqno(45)
$$
with $\delta\xi$, which is equal to the sum of the RHS's of Eqns. (43a-b),
providing a new fluctuation source beside the original term $\hat\Delta\xi$.

At this point we are in the condition to identify the terms in the shell
energy equation which are responsible for the generation of intermittency.
For the sake of simplicity, and to make contact with the model of Eggers
$[22]$, let us adopt for a moment the approximation: $\xi_n\simeq
e^{-\gamma n/2}\hat\xi$ with $<\hat\xi(t)\hat\xi(t')>=\delta(t-t')$, and
similar equation for $\delta\xi$. The RHS of Eqn. (45) is then proportional to:
$$
\hat\Delta\hat\xi-\gamma\hat\xi/2+\delta\hat\xi.\eqno(46)
$$
We see then that Eqn. (45), at statistical equilibrium, is very similar to a
random walk equation in which the role of the time is played by $n/V$ and that
of the random kicking by $-\gamma\hat\xi/2+\delta\hat\xi$. This noise term
continuously pumps into the system fluctuations, which are dissipated at very
large $n$, by viscosity. It is the random walk character of the
process that leads to the linear growth of $<\phi_n^2>$, with respect to $n$,
already observed in $[22]$. The two terms providing the source of intermittency
have different physical origin. The term $\delta\xi$, which comes from Eqns.
(44a,b), is due to the competition in the energy transfer between eddies
at different locations, characteristic of the Random Beta Model $[8]$.
The term $\gamma\hat\xi/2$ instead, comes from the mismatch in the
characteristic time scales of the energy transfer between different shells and
was responsible for the production of intermittency in the stochastic model of
Eggers $[22$-$23]$.

The first term in Eqn. (46), which is the derivative of a random noise,
produces the gaussian part of the fluctuations in the energy content of
the wavepackets. Notice that this same quantity can be calculated directly
from finite volume averages of $|\v^\smalze|^2$, given the expression for the
correlation given by Eqn. (3), together with Eqn. (18). This fact will be used
to provide a check on the goodness of the approximations used
to arrive to equation (45). Notice finally how in this approach, the smallness
of the anomalous corrections is associated with the smallness of the parameter
$a-1$ and with the fact that the amplitude of the intermittency source term is
second order in this quantity, with respect to the source term of the gaussian
fluctuations.
\vskip 10pt

\centerline{\bf A. Solution of the shell energy equation}
\vskip 5pt
It is possible to solve Eqn. (45) either following $[22]$, by diagonalizing
the LHS of (45), considered as a matrix equation, or using the multiplier
technique adopted in $[23]$. Here, the smallness of the parameter $a-1$ allows
to consider the continuous limit of Eqn. (45) and to use a multiple scale
expansion in which, to lowest order, the dependence on $n$ produced by the
$\exp(\gamma n)$ terms is neglected on the scale of the fluctuations. This
allows a solution of the problem in terms of green functions, in which no
approximation on the form of the noise correlation is required.
The green function for Eqn. (45) is:
$$
g(n,n',t)=
{\exp\Big[-{(n-n'-V\,e^{\gamma n}t)^2\over 4D\,e^{\gamma n}t}
+{\gamma n\over 2}\Big]\over\sqrt{4\pi D\,t}},\eqno(47)
$$
The first quantity that we are going to calculate is the gaussian part of the
fluctuations:
$$
<\phi_G^2>=F\int{\d k\over 2\pi}{\d\omega\over 2\pi}|g_{k\omega}(n)|^2k^2\,
<|\xi_{k\omega}|^2>
$$
$$
={F\over 2D}\int{\d k\over 2\pi}{(1+Dk^2)\exp(-{0.14\over a-1}|k|)\over
1+(V^2+2D)k^2+D^2k^4},\eqno(48)
$$
where $g_{k\omega}(n)=(-i(\omega\,e^{-\gamma n}-Vk)+Dk^2)^{-1}$ is the
Fourier transform with respect to $n'$ and $t$ of $g(n,n',t)$.
As mentioned before, we can repeat the calculation directly from the statistics
of the velocity field $\v^\smalze$. Considering the limit of $c_w$ small,
corresponding to wavepackets thinner than the shell, nondiagonal contributions
in Eqn. (18) can be neglected and we have:
$$
<\phi_G^2>={2\over E_n^{\smalze 2}}\int_n{\d^3 k\over (2\pi)^3}C_k^2
\int_n{\d^3 q\over (2\pi)^3}|W_q(n)|^2
$$
$$
\simeq 0.032{(1-a^{-{13\over3}})(c_w(a-1))^3\over(1-a^{-{2\over 3}})^2}.
\eqno(49)
$$
The two expressions for $<\phi_G^2>$ given by  Eqns. (48) and (49) are plotted
in Fig. 5. against $a$; the best agreement, though still rather rough, is
obtained for the range $1.2\le a\le 1.3$, which is consistent with what would
be
expected by looking at the energy transfer profile (Fig. 2.) and at the one for
the correlation of the energy flux fluctuations (Fig. 3.).
In the same way it is possible to calculate the correlation time for $\phi_G$:
$$
\tau_c={1\over<\phi_G^2>}\int_0^\infty\d t\,<\phi_G(0)\phi_G(t)>
$$
$$
={-iF\over<\phi_G^2>}\int{\d k\over 2\pi}{\d\omega\over 2\pi}
{k^2\over \omega}|g_{k\omega}(n)|^2<|\xi_{k\omega}|^2>.\eqno(50)
$$
The prediction from direct calculation, obtained from the generalization of
Eqn. (49) to 2-time correlations, is $\tau_c(k_0)\simeq 1$, corresponding,
in non rescaled units, to $\tau_c(k)^{-1}=2\eta_{k}$; here we find that
$0.9\le\tau_c(k_0)\le 1$ in the whole range $1.1\le a\le 2$.

\vskip 5pt
We turn next to the calculation of the intermittent part of the fluctuation
$\phi$. Care must be taken now due to the divergent nature of correlations at
large $n$, which forbids in particular the use of the Fourier representation
adopted in Eqn. (48).

The equation giving the growth of $<\phi_n^2>$ at large $n$ is obtained by
multiplying Eqn. (45) by $\phi$ and taking the average:
$$
<\phi_n[\exp(-\gamma n)\partial_t-D\Delta^2+V\Delta]\phi_n>=
F^{1/2}<\phi_n(\hat\Delta\xi_n+\delta\xi)>.\eqno(51)
$$
We subtract from Eqn. (51) the gaussian part of the fluctuations as given by
Eqn. (48); then the
finite difference $\hat\Delta$ acts only on the part of the noise variation
due to the scaling of the correlation time, which is of order $a-1$. Next, we
notice that to lowest order, the contribution to $\delta\xi$ coming from
Eqn. (43a) is obtained by approximating $\phi$ by its gaussian component
$\phi_G$. The RHS of Eqn. (51) takes then the form:
$$
\Xi(a,c_w)=F\int_0^\infty\d t\int\d n'\,[g(n,n',t)\,(\gamma^2(1-t^2)+\beta)
$$
$$
+g(n,n'-1,t)\beta't]\,<\xi_n(0)\xi_{n+n'}(t)>
+\beta<\phi_G(D-{V\over 2})\phi_G>,\eqno(52)
$$
where $\beta$ and $\beta'$ are $\O((a-1)^2)$ quantities giving respectively
the amount of
wavepacket volumes not overlapping (Beta model effect), and the correlation
between this effect and that of the scale dependence of the noise correlation
time. Explicit expressions for these coefficients and further simplification
of Eqn. (52) are presented in Appendix C.

Next turn to the LHS of Eqn. (51). The time derivative term is equal to zero
at steady state, while it is shown in Appendix C that the intermittent part of
the fluctuations does not contribute to the $<\phi\Delta^2\phi>$ term. We
are left then with: $<\phi\Delta\phi>\simeq\unmezzo\Delta<\phi^2>$, so that
we obtain the result, for the kurtosis scaling exponent:
$$
2\zeta_2-\zeta_4\simeq{1\over\ln a}{\d<\phi^2>\over \d n}={2\,\Xi(a,c_w)\over
V(a)\,\ln a}.\eqno(53)
$$
The dependence of $2\zeta_2-\zeta_4$ on $a$ for fixed $c_w$ is shown in Fig.
6.;
notice the saturation occurring at $a\simeq 1.3$, suggesting that the bulk of
intermittency production occurs at scales of the order of three to ten times
the
size of the eddies in exam. From inspection of Eqns. (43), (52), (53) and
(C3-6),
we see that, for small $a-1$: $2\zeta_2-\zeta_4=\O(c_w^3(a-1)^3)$, so that we
obtain a direct connection between the smallness of the intermittency
corrections and their being proportional to a rather large power of the
small parameter $a-1$. Unfortunately, the intermittency exponents defined here
depend sensitively on the parameter $c_w$; due to the difficulty in determining
$c_w^\smalMAX$ with precisions, it is therefore problematic to
make accurate predictions on the scaling of $S(l,q)$ starting from
$S(l,a,c_w,q)$.

A value $2\zeta_2-\zeta_4\simeq 0.016$, is obtained setting $c_w=2$,
corresponding to the ''reasonable'' condition that the noise granularity be
equal to the shell thickness $(a-1)k_n$, i.e., from Eqns. (19) and (29):
$W_{\p+\q-\k}^2\sim\exp(-{|\p+\q-\k|^2\over 2(k_{n+1}-k_n)^2})$. However
choices
as reasonable as the one just considered, lead to results differing from one
another by up to an order of magnitude, so that the above formula should not
be taken too seriously.

\vskip 20pt
\centerline{\bfm V. Summary and conclusions}
\vskip 5pt
We have described a model for the production of intermittency in the inertial
range of three dimensional turbulence, based on statistical closure of the
Navier Stokes equation. A connection between Navier Stokes dynamics and
phenomenological models like the Random Beta Model $[8]$, and the stochastic
chains considered in $[22$-$23]$, is in this way established. Although this
connection may be rather tenuous, because of the assumptions adopted in
deriving
the closure, it is still pleasing that the results in this paper are obtained
as lowest order corrections to a mean field approximation, which by itself
would produce Kolmogorov scaling. It should also be mentioned that the present
approach is able to produce dynamically, values of non-intermittent part of the
energy fluctuations in agreement with the prediction from quasigaussian
statistics of the velocity field; this is a bonus, which provides an indirect
check on the goodness of the closure technique adopted.

The use of wavepackets rather than Fourier modes, is the reason why a
perturbative treatment of intermittency has been possible here.
There has been indeed an intriguing
aspect in this subject, namely, the difficulty in associating, to the smallness
of the anomalous corrections, a small parameter in which to carry on
perturbation theory. One of the results of the present model is the
identification of this parameter with the physical quantity $\Delta k/k$, i.e.
the ratio between eddy size and the scale of the characteristic flow straining
the eddy. Scaling corrections appear to be of third order in this quantity,
with
equally important contributions from Random Beta Model kind of effects $[8]$,
and from the mechanism of intermittency production of the model studied by
Eggers $[22$-$23]$.

The main result of this paper justifies a posteriori the use of wavepackets
in the problem: the saturation in the $a$-dependence of the generalized
structure functions $S(l,a,c_w,q)$, for $a\ge 1.3$, which is consistent
with a separation of scales between production of intermittency and energy
transfer. This is a definite prediction of the model, which, together with
predictions on the actual magnitude of scaling corrections, could be tested by
direct analysis of experimental data in terms of the generalized structure
functions $S(l,a,c_w,q)$. This would extend similar studies, carried on by
Menevau $[30]$ using wavelet analysis.

We have not been able yet to establish a quantitatively accurate relation
between the two structure function definitions: $S(l,a,c_w,q)$ and $S(l,q)$,
the reason
being the sensitive dependence of $S(l,a,c_w,q)$ on $c_w={\Delta
k\over(a-1)k}$,
the ratio of wavepacket to shell thickness in $k$-space. Preliminary estimates
hint towards a value for the scaling corrections smaller by a factor
of the order of five than the experimental one. In the lognormal approximation:
${\delta\zeta_q^{\scriptscriptstyle theo.}\over q(q-3)}\sim-.002$, which
should be compared with the lognormal fit of experimental values:
${\delta\zeta_q\over q(q-3)}\sim-.01$ $[2]$. Due to uncertainties in the
present
analysis, it is not possible yet to state whether inertial range processes are
as important for intermittency production as finite size corrections, or if
they
are themselves just corrections to dominant finite size effects. We reiterate
however that such an answer could be obtained analyzing experimental data in
terms of $S(l,a,c_w,q)$ instead of $S(l,q)$.

Turning our attention to more formal issues, it is interesting to look for
similarities between the model described in this paper and the various
phenomenological approaches that have been used to study infinite Reynolds
number intermittency. The basic ingredient here is the partition of
Fourier space in shells of exponentially increasing radii. For this reason,
our approach  has a lot in common with the deterministic Shell Models studied,
among others, by Yamada and Ohkitani $[31]$ and by Jensen, Paladin and Vulpiani
$[32]$. An interesting interpretation of these models, particularly clear in
Zimin $[33]$, comes from looking at the dynamical variable in each shell as the
velocity of nested eddies, all located at the same space position, and
interacting with one another locally in both scale and space. In our approach,
such dramatic restriction of phase space does not take place, in that the
behavior of the various wavepackets in a given shell is treated in an average
sense; it is then possible that our model may underestimate the amount of
intermittency produced in the inertial range. It remains to be seen how
important this difference is; in principle one should compare the
present model with some closure for the Yamada-Ohkitani model or, viceversa,
look for a deterministic dynamical system, whose closure coincide with the
one described in this paper, and then compare with the Yamada-Okhitani
system.

The connection with the stochastic model of Eggers $[22$-$23]$ is clearly
simpler. In both cases one ends up with the same stochastic chain, the only
difference being here the not exact conservation of energy transferred between
shells. If this effect, due to the not exact overlapping of wavepackets at
different scales, were neglected, the present model and the one of Eggers would
coincide.

Going back to the possibility of turbulent Navier Stokes dynamics taking place
in a very limited region of phase space, this constitutes the main conceptual
limitation of our model. The possibility of preferential transfer of energy
between individual wavepackets would correspond to intermittency being
associated with locally anisotropic fluctuations. It should be mentioned
that this looks somewhat unlikely, due to the separation of scales
described earlier and the consequent large number of eddies in one straining
region $\delta k^{-1}$. In any case this is a possibility, which could lead,
in a suggestive way, to a mechanism for the creation of coherent structures.
In $[34]$, She derived anomalous exponents, based on assumptions on the shape
of the most intense and rare dissipation regions, so that intermittency and
coherent structures may be connected. It is difficult to see how
shell models, of either the deterministic or the stochastic variety, could be
used to study such coherent structures, given the very distant scales involved
in their dynamics. In any case, perhaps this is not a big problen, since,
although vortex tubes have been observed in many numerical simulations $[3]$,
they do not seem to contribute much to energy dissipation and their relevance
to the turbulent dynamics has been questioned recently in $[35]$.

\vskip 20pt
\noindent{\bf Aknowledgements:} I would like to thank Jens Eggers, Gregory
Falkovich and Detlef Lohse for interesting and very helpful discussion. This
research was supported in part by DOE, the ONR and the University of Chicago
MRL.
\vfill\eject

\centerline {\bf References}
\vskip 5pt
\item{$[1]$} A.N. Kolmogorov, ''Local structure of turbulence in an
incompressible fluid at very large Reynolds numbers,'' CR. Acad. Sci. USSR.
{\bf 30},299 (1941)
\item{$[2]$} F. Anselmet, Y. Gagne, E.J. Hopfinger, and R. Antonia, ''High
order velocity structure functions in turbulent shear flows,'' J. Fluid Mech.
{\bf 140}, 63 (1984)
\item{$[3]$} A. Vincent and M. Meneguzzi, ''The spatial structure and
statistical properties of homogeneous turbulence,'' J. Fluid Mech. {\bf 225}, 1
(1991)
\item{$[4]$} R. Benzi, S. Ciliberto, R. Tripiccione, C. Baudet, F. Massaioli
and
S. Succi, ''Extended self-similarity in turbulent flows,''
Phys. Rev. E {\bf 48} R29 (1993)
\item{$[5]$} L.D. Landau and E.M. Lifsits, {\it Fluid mechanics}, (Pergamon,
Oxford, 1984)
\item{$[6]$} A.N. Kolmogorov, ''A refinement of previous hypothesis concerning
the local structure of turbulence in a viscous incompressible fluid at high
Reynolds numbers,'' J. Fluid Mech. {\bf 13}, 82 (1962)
\item{$[7]$} G. Parisi and U. Frisch, ''On the singularity structure of fully
developed turbulence'' in {\it Turbulence and predictability of geophysical
fluid dynamics}, edited by M. Ghil , R. Benzi and G. Parisi (North Holland,
Amsterdam, 1985), p. 84
\item{$[8]$} R. Benzi, G. Paladin, G. Parisi and A. Vulpiani, ''On the
multifractal nature of fully developed turbulence and chaotic systems,''
J. Phys. A {\bf 18}, 2157 (1985)
\item{$[9]$} B.B Mandelbrot, ''Intermittent turbulence in self-similar
cascades:
divergence of high moments and dimensions of the carrier,'' J. Fluid Mech.
{\bf 62}, 331 (1974)
\item{$[10]$} T.C. Halsey, M.H. Jensen, L.P. Kadanoff, I. Procaccia and B.I.
Shraiman, ''Fractal measures and their sincularities: The characterization of
strange sets,'' Phys. Rev. A {\bf 33}, 1141 (1986)
\item{$[11]$} S. Grossmann and D. Lohse, ''Scale resolved intermittency in
turbulence,'' Phys. Fluids A {\bf 6}, 711 (1994)
\item{ } D. Lohse and A. M\"uller Groeling, ''Bottleneck effects in turbulence:
Scaling
phenomena in $r$-versus $p$-space, preprint, Chicago and Toronto (1994)
\item{$[12]$} V.V. Lebedev and V.S. L'vov, ''Scaling of correlation functions
of
velocity gradients in hydrodynamics turbulence,'' JETP Letters {\bf 59}, 577
(1994)
\item{ } V.S. L'vov, I. Procaccia and A.L. Fairhall, ''Anomalous scaling in
fluid
mechanics: the case of the passive scalar'', (1994) Phys. Rev. E, {\it in
press}
\item{$[13]$} U. Frisch and R. Morf, ''Intermittency in nonlinear dynamics and
singularities at complec times,'' Phys. Rev. A {\bf 23}, 2673 (1981)
\item{$[14]$} V. Yakhot, ''Spectra of velocity, kinetic energy, and the
dissipation rate in strong turbulence,'' Phys Rev. E {\bf 50}, R20 (1994)
\item{$[15]$} {\it Indeed, quantities in the form $<f(x)^q>$, correspond in
Fourier space to convolutions. This has the important consequence that real
space intermittency is not associated with intermittency of the Fourier modes,
but rather with their being strongly correlated. In particular, a situation in
which the quantity $|f_k|$ is intermittent, but the correlations between modes
obey gaussian statistics, can be shown to lead to a non intermittent $f(x)$.}
\item{$[16]$} H. Effinger and S. Grossmann, ''Static structure function of
turbulent flow from the Navier Stokes equations,'' Z. Phys. B {\bf 66}, 289
(1987)
\item{$[17]$} T. Nakano, ''Direct interaction approximation of turbulence in
the
wavepacket representation,'' Phys. Fluids {\bf 31}, 1420 (1988)
\item{$[18]$} J. Eggers and S. Grossmann, ''Anomalous turbulent scaling from
the
Navier Stokes equation,'' Phys. Lett. A {\bf 156}, 44 (1991)
\item{$[19]$} J.A. Domaradzki and R.S. Rogallo, ''Local energy transfer and
nonlocal interaction in homogeneous, isotropic turbulence,'' Phys Fluids A
{\bf 2}, 413 (1990)
\item{$[20]$} S.A. Orszag ''Lectures on the statistical theory of turbulence,''
in {\it Fluid Dynamics}, edited by R. Balian and J.-L. Peube (Gordon and
Breach,
New York, 1977), p. 235
\item{$[21]$} V.S. L'Vov, ''Scale invariant theory of fully developed
hydrodynamic turbulence; Hamiltonian approach,'' Phys. Rep. {\bf 207}, 1 (1991)
\item{$[22]$} J. Eggers, ''Intermittency in dynamical models of turbulence,''
Phys. Rev. A {\bf 46}, 1951 (1992)
\item{$[23]$} J. Eggers, ''Multifractal scaling from nonlinear turbulence
dynamics: Analytical methods,'' Phys Rev. E {\bf 50}, 285 (1994)
\item{$[24]$} R.H. Kraichnan, ''The structure of isotropic turbulence at very
high Reynolds number,'' J. Fluid Mech. {\bf 5}, 497 (1959)
\item{$[25]$} R.H. Kraichnan, ''Lagrangean history closure approximation for
turbulence,'' Phys. Fluids {\bf 8}, 575 (1967)
\item{$[26]$} D.C. Leslie, {\it Developments in the theory of turbulence},
(Clarendon Press, Oxford, 1973)
\item{$[27]$} R.H. Kraichnan, ''Inertial range transfer in two and three
dimensional turbulence,'' J. Fluid Mech. {\bf 47}, 525 (1971)
\item{$[28]$} {\it Notice that the shift to a Lagrangean frame does not
introduce cutoffs in the integrals at small $pq$ neither in Eqn. (11) nor in
(15); what happens is that while in the Eulerian DIA the green function
expression is associated with the sink term $C_qC_k$ in the energy equation,
here, the new terms generated in the Lagrangean reference frame, are associated
with contributions to the source $C_pC_q$ in (15).}
\item{$[29]$} V. Zakharov, V. L'vov and G. Falkovich, {\it Kolmogorov spectra
of turbulence} (Springer, Heidelberg, 1992)
\item{$[30]$} C. Menevau, ''Analysis of turbulence in the orthonormal wavelet
representation'', J. Fluid Mech. {\bf 232}, 469 (1991)
\item{$[31]$}  M. Yamada and K. Ohkitani, ''The constant of motion and the
inertial subrange spectrum in fully developed model turbulence,'' Phys. Lett. A
{\bf 134}, 165 (1988)
\item{$[32]$} M.H. Jensen, G. Paladin and A. Vulpiani, ''Intermittency in a
cascade model for three-dimensional turbulence,'' Phys. Rev. A {\bf 43}, 798
(1991)
\item{$[33]$} V.D. Zimin, ''Hierarchic model of turbulence,'' Atmospheric and
Oceanic Physics, {\bf 17}, 941 (1981)
\item{$[34]$} Z.S. She and E. Leveque, ''Universal scaling laws in fully
developed turbulence,'' Phys. Rev. lett. {\bf 72}, 336 (1994)
\item{$[35]$} J. Jim\`enez, A.A. Wray, P.G. Saffman and R.S. Rogallo, ''The
structure of intense vorticity in isotropic turbulence,'' J. Fluid Mech. {\bf
255}, 65 (1993)
\item{$[36]$} I.S. Gradhsteyn and I.M. Ryzhik {\it Table of integrals, series
and products}, (Academic Press, San Diego, 1980)
\vfill\eject

\noindent{\bf Appendix A. Quasi Lagrangean closure}
\vskip 5pt
The basic difficulty in dealing with Quasi Lagrangean closures is that studying
the Navier Stokes dynamics in a reference frame moving with a single Lagrangean
tracer, neglects the divergence of trajectories of different tracers. In a more
refined closure scheme, this effect would be included by associating each point
in a correlation function to a different tracer trajectory, like in the
Lagrangean History DIA (LHDIA) of Kraichnan $[25]$. However, also in that
theory,
abridgements of the closure equations were necessary in the end, which were
similar in their effect to neglecting the divergence of Lagrangean
trajectories.

We try here to implement this approximation in as much a consistent way as
possible. Notice first that correlations are defined starting from an initial
case in which the initial point lies on the Lagrangean trajectory $\zr_t$, so
that Eqn. (3) basically describes the decorrelation of points at distance
$r$ from the Lagrangean trajectory $\zr_t$, with respect to points lying on it.
Since decorrelation occurs forward in time, this is the optimal
choice; choosing $\zr_t$ as the final point and $\zr_0+\r$ as the initial one,
would
lead in particular to no decorrelation, due to the advection term being zero in
Eqn. (2).

The approximation of Eqn. (5) in which the initial point $\zr_{t_1}+\r_1$ of
the
green function $G(t,\r |t_1,\r_1)$ is shifted on the Lagrangean trajectory, is
justified with the previous choice. We try to make it more appealing by
giving the next order in the expansion around $G(t,\r |t_1,\r_1)
=G(\r-\r_1,t-t_1)$. We have first:
$$
\int\d^3 r\,e^{-i\k\r}<v^\alpha(\r_0,0)v^\gamma(\zr_t(\r_0,0)+\r,t)>=
\int\d^3 r\,e^{-i\k\r}\int\d^3 s\,G^{\gamma\rho}(0,\r+\ss |t,\r)
C^\alpha_\rho(\r-\ss)
$$
$$
=\int\d^3 r\int{\d^3 k_1\over (2\pi)^3}G^{\gamma\rho}_{\k_1}(\r,t)
C_{\k_1,\rho}^\alpha\,e^{i(\k_1-\k)\r},\eqno({\rm A}1)
$$
where $G^{\gamma\rho}_{\k_1}(\r,t)$ is the Fourier transform of
$G^{\gamma\rho}(0,\r+\ss |t,\r)$ with respect to $\ss$. Expanding the argument
of the integral in the last line of Eqn. (A1) in $\k-\k_1$ and $\r$, we get
then:
$$
\int\d^3 r\,e^{-i\k\r}<v^\alpha(\r_0,0)v^\gamma(\zr_t(\r_0,0)+\r,t)>
$$
$$
=G^{\gamma\rho}_\k(0,t)C^\alpha_{\k,\rho}-\unmezzo{\partial\over\partial
k^\phi}
{\partial\over\partial k^\psi}{\partial\over\partial r_\phi}{\partial\over
\partial r_\psi}\Big[G_\k^{\gamma\rho}(\r,t)C_{\k,\rho}^\alpha\Big]_{\r=0}+...
\eqno({\rm A}2)
$$
More in general we would obtain:
$$
\int\d^3 r\,e^{-i\k\r}<v^\alpha(\zr_{t_1}(\r_0,0)+\r_1,t_1))
v^\gamma(\zr_{t_1+t}(\r_0,0)+\r_1+\r,t_1+t)>
$$
$$
=G^{\gamma\rho}_\k(\r_1,t)C^\alpha_{\k,\rho}-\unmezzo{\partial\over\partial
k^\phi}
{\partial\over\partial k^\psi}{\partial\over\partial r_\phi}{\partial\over
\partial r_\psi}\Big[G_\k^{\gamma\rho}(\r_1+\r,t)C_{\k,\rho}^\alpha\Big]_{\r=0}
+...
\eqno({\rm A}3)
$$
which can be then be expanded in Taylor series around $\r_1=0$. This shows that
the lowest order approximation adopted in Eqn. (5) is rather bad and that our
closure could not be expected to lead to quantitative accurate results.

\vskip 5pt
We turn next to the derivation of the energy balance equation in the laboratory
frame, Eqn. (10). We consider just one term:
$$
-{1\over (2\pi)^3}\int\d^3r_1\d^3r_2\d^3r_3\,e^{-i(\k\r_1-\p\r_2-\q\r_3)}
<v^{\smalze\alpha}(\r_1,0)v^\smalze_\beta(\r_2,0)\partial^\beta
v^\smalun_\alpha(\r_3,0)>
$$
$$
={1\over (2\pi)^3}\int\d^3r_1\d^3r_2\d^3r_3\,e^{-i(\k\r_1-\p\r_2-\q\r_3)}
\int\d^3 s\int_{-\infty}^0\d\tau\,G_\alpha^\rho(\ss,-\tau)
$$
$$
\times<v^\alpha(\r_1,0)\v_\beta(\r_2,0)\partial^\beta
[v^\sigma(\zr_\tau+\r_3+\ss,\tau)-\hat wv^\sigma(\zr_\tau,\tau)]\partial_\sigma
v_\rho(\zr_\tau+\r_3+\ss,\tau)>.\eqno({\rm A}4)
$$
Splitting the 4-point correlation into 2-point ones, we obtain, in terms of
Fourier transforms:
$$
2\delta(\k-\p-\q)\int\d^3r_2\d^3r_3
\int\d^3k_1\d^3k_2\d^3k_3\delta(\k+\p+\k_2-\k_3)\theta_{k_1k_2k_3}C_{k_2}C_{k_3}
$$
$$
\times\Big\{B_4(k_1k_2k_3)
[\delta(\k_1-\k_2-\k_3)\delta(\k+\q+2\k_2+\k_3)-
w\delta(\k_1-\k_3)\delta(\k+\q+\k_2+\k_3)]
$$
$$
+B_2(k_1k_2k_3)[\delta(\k_1-\k_2-\k_3)\delta(\k+q+2\k_2+\k_3)-
w\delta(\k_1-\k_2)\delta(\k+\q+2\k_2)]\Big\},\eqno({\rm A}5)
$$
where:
$$
B_2(k_1k_2k_3)=P^{\alpha\rho}(k_1)P_{\alpha\rho}(k_2)k_1^\beta
P_{\beta\sigma}(k_3)k_2^\sigma
=k_1k_2(1+z^2)(z+xy),\eqno({\rm A}6)
$$
and
$$
B_4(k_1k_2k_3)=k_2^\sigma P_{\sigma\alpha}(k_1)P^{\alpha\rho}(k_2)
P_{\rho\beta}(\k_3)k_1^\beta=-k_1k_2xz(x+yz).\eqno({\rm A}7)
$$
The various terms in Eqn. (A5) are tracked back to Eqn. (A4) as follows: the
terms in $B_4$ come from the contraction $<v^\alpha(v^\sigma-wv^\sigma)>
\partial^\beta G_\alpha^\rho<v_\beta\partial_\sigma v_\rho>$; those in $B_2$,
from
the remaining contraction; the terms in $w$ come from the shift to Lagrangean
frame. Notice now that, from Eqns. (A6-7): $B_2(k_1k_1k_3)=B_4(k_1k_2k_1)=0$.
Thus the terms in $w$ in Eqn. (A5) disappear and one is left with the same
expression that would be obtained from Eulerian DIA $[24]$, but with the
Lagrangean response time $\theta$. Repeating the same calculation with the
other two choices for $v^\smalun$, we see that no terms in $w$ contribute and
we obtain the standard result of Eqn. (10).

\vskip 5pt
We pass to the calculation of the 2-time Lagrangean correlation: at steady
state
the part of the time integrals from $\tau<0$ do not contribute and we have:
$$
-\int\d^3 r\,e^{-i\k\r}<v^\alpha(\r_0,0)[v^\beta(\zr_t+\r,t)-\hat w
v^\beta(\zr_t,t)]\partial_\beta v_\alpha(\zr_t+\r,t)>
$$
$$
=\int\d^3 r e^{-i\k\r}\int_0^t\d\tau\int\d^3 s\Big\{G^\beta_\rho(\ss,t-\tau)
\Big[[C^{\alpha\sigma}(-\r-\ss,\tau)-\hat wC^{\alpha\sigma}(0,\tau)]
\partial_\sigma\partial_\beta C^\rho_\alpha(-\ss,t-\tau)
$$
$$
+\partial_\beta[C_\alpha^\sigma(-\ss,t-\tau)-\hat w C_\alpha^\sigma(\r,t-\tau)]
\partial_\sigma C^{\rho\alpha}(-\r-\ss,\tau)\Big]
$$
$$
-\hat wG^\beta_\rho(\ss,t-\tau)\Big[[C^{\alpha\sigma}(-\ss,\tau)-\hat w
C^{\alpha\sigma}(0,\tau)]\partial_\sigma\partial_\beta
C^\rho_\alpha(\r-\ss,t-\tau)
$$
$$
+\partial_\beta[C_\alpha^\sigma(\r-\ss,t-\tau)-\hat w
C_\alpha^\sigma(\r,t-\tau)]\partial_\sigma C^{\rho\alpha}(-\ss,\tau)\Big]
$$
$$
+\partial_\beta G_{\alpha\rho}\Big[[C^{\alpha\sigma}(-\r-\ss,\tau)-
\hat wC^{\alpha\sigma}(0,\tau)]
$$
$$
\times\partial_\sigma[C^\beta_\rho(-\ss,t-\tau)-
\hat C^\beta_\rho(-\r-\ss,t-\tau)]
$$
$$
+\partial_\sigma C^{\alpha\rho}(-\r-\ss,\tau)
[C^{\beta\sigma}(-\ss,t-\tau)-\hat wC^{\beta\sigma}(\r,t-\tau)
$$
$$
-\hat wC^{\beta\sigma}(-\r-\ss,t-\tau)+\hat w\hat wC^{\beta\sigma}(0,t-\tau)]
\Big]\Big\}.\eqno({\rm A}8)
$$
Of the terms on the RHS of Eqn. (A8), there is a group which does not depend
on the integration variable $\ss$; these terms come from integrating along
the Lagrangean trajectory and lead, after Fourier transform, to triads in
which one of the wavevectors is zero; as before they do not contribute to
the final result. The remaining terms in $\hat w$ come from the
$\hat w$ on the LHS of the equation, which express the fact that we are dealing
with a Lagrangean correlation. These terms remain and are responsible for
the cancellation of the sweep part of the correlation decay. After Fourier
transorm, we obtain the result:
$$
\int_0^t\d\tau\int{\d^3 p\over (2\pi)^3}{\d^3 q\over (2\pi)^3}\delta(\k-\p-\q)
$$
$$
\times\Big[B_3(kpq)G_p(t-\tau)C_k(\tau)C_q(t-\tau)
-w_pB_3(q,-p,k)G_p(t-\tau)C_k(t-\tau)C_q(\tau)]
$$
$$
+[B_1(kpq)G_p(t-\tau)C_k(\tau)C_q(t-\tau)
-w_pB_1(q,-p,k)G_p(t-\tau)C_k(t-\tau)C_q(\tau)]
$$
$$
+[B_4(kpq)G_p(t-\tau)C_k(\tau)C_q(t-\tau)
-w_qB_4(p,k,-q)G_k(t-\tau)C_p(\tau)C_q(t-\tau)]
$$
$$
+B_2(kpq)G_p(t-\tau)C_k(\tau)C_q(t-\tau)
-w_qB_2(p,k,-q)G_k(t-\tau)C_p(\tau)C_q(t-\tau)]\Big],\eqno({\rm A}9)
$$
where:
$$
B_1(kpq)=p_\sigma P^{\sigma\alpha}(q)P_{\alpha\rho}(k)P^{\rho\beta}(p)k_\beta
=-kp\,yz(y+xz)\eqno({\rm A}10)
$$
and
$$
B_3(kpq)=k_\beta P^{\beta\rho}(p)P_{\rho\alpha}(q)P^{\alpha\sigma}(k)p_\sigma
=kp\,xy(1-z^2).\eqno({\rm A}11)
$$
(The same notation of Leslie $[26]$ is used here for the functions $B_i$,
$i=1,...4$). The terms on the RHS of Eqn. (A9) are ordered as on the RHS of
Eqn. (A8), once the terms of the last one, that are equal to zero, are
eliminated. Using the following relations:
$$
B_3(q,-p,k)=-B_1(kpq);\qquad B_1(q,-p,k)=-B_3(kpq)
$$
$$
B_4(p,k,-q)=B_1(kpq);\qquad B_2(p,k,-q)=B_2(kpq),\eqno({\rm A}12)
$$
and the definitions:
$$
b_{kpq}={1\over 2 k^2}\sum_i B_i(kpq);\qquad
b^\smalun_{kpq}={1\over 2k^2}(B_1(kpq)+B_3(kpq));
$$
$$
b^\smaldu_{kpq}={1\over 2k^2}(B_1(kpq)+B_2(kpq)),\eqno({\rm A}12)
$$
and substituting into Eqn. (A9), we obtain immediately the result of Eqn. (11).

\vskip 5pt
Finally, we derive the energy equation in the moving reference frame. Using
conservation of energy triad by triad (which is preserved, together with
incompressibility, when passing to the Lagrangean reference frame), we can
write:
$$
D_t C_k(t)|_{t=0}={1\over 4\pi}\int_\Delta\d p\d q\,kpq\theta_{kpq}
[-H(kpq)+H(pkq)+H(qkp)],\eqno({\rm A}13)
$$
where:
$$
H(kpq)=[b_{kpq}+w_p\bun]C_kC_q-w_q\bdu C_pC_q.
\eqno({\rm A}14)
$$
We obtain immediately:
$$
D_t C_k(t)|_{t=0}={1\over 4\pi}\int_\Delta\d p\d q\,kpq\theta_{kpq}
[(b_{kpq}+w_q\btr)C_pC_q
$$
$$
-(b_{kpq}+({p\over k})^2w_qb^\smaltr_{p,k,-q})C_kC_q],\eqno({\rm A}14)
$$
where:
$$
\btr={1\over 2k^2}(2B_2(kpq)+B_1(kpq)+B_4(kpq)).\eqno({\rm A}15)
$$
A form of Eqn. (A14), which is  manifestly zero at equipartition, can be
obtained using the relations:
$$
B_2(kpq)=B_2(p,k,-q)\qquad{\rm and}\qquad B_1(kpq)=B_4(p,k,-q).\eqno({\rm A}16)
$$
Substituting Eqns. (A15-16) into Eqn. (A14), we obtain finally the result shown
in Eqn. (15).

\vskip 20pt
\noindent{\bf Appendix B. Noise correlation}
\vskip 5pt
In order to calculate the amplitude of the energy flux fluctuation $g_n(m)$,
we must evaluate the contractions of the product:
$$
v^\alpha(\zr_1)[v^\beta(\zr_1+\r_1)-v^\beta(\zr_1)]\partial_\beta
v^\alpha(\zr_1+\r_1)v^\gamma(\zr_2)[v^\sigma(\zr_2+\r_2)-v^\sigma(\zr_2)]
\partial_\sigma v^\gamma(\zr_2 +\r_2),\eqno({\rm B}1)
$$
which are listed in Table B1. In this section, we shall adopt the notation:
$\psi_\alpha\equiv\partial_\alpha\psi$. Of these contractions, numbers (B1.1),
(B1.4),
(B1.7), (B1.10) and (B1.13) do not contribute because of incompressibility.
The remaining ones fall into two groups: (B1.2), (B1.3), (B1.4) and (B1.6)
are in the form: $<g(0)g(t)>\to<v(0)v(0)><v(0)v(t)><v(t)v(t)>$, where
$<v(0)v(t)>$ is associated with velocity fluctuations on the scale of the
wavepacket radius $R$. These terms come from noise components of $g$ in the
form $v<(v\nabla)v>$, which are identicaly zero, and $(v\nabla)<v^2/2>$, which
are associated with sweep and do not contribute to energy transfer.
The second term contains all the other terms, which are in the form
$<v(0)v(t)><v(0)v(t)><v(0)v(t)>$ and act on the same scale of $k_1$ and $k_2$.

It is possible to show explicitly that the contractions in the first group do
not contribute to $<g^2>$. Consider for example contraction (B1.3):
$$
<g_{n_1}(m_1,t_1)g_{n_2}(m_2,t_2)>^\smaltr=
\int\d^3 r_1\d^3 r_2H(r_1,n_1)H(r_2,n_2)
$$
$$
\times\int\d^3 z_1\d^3 z_2W(z_1,m_1)W(z_2,m_2)
[C^{\alpha\beta}(\r_1,0)-\hat wC^{\alpha\beta}(0,0)]
$$
$$
\times[C^{\alpha\sigma}_\beta(\zr_1+\r_1-\zr_2-\r_2,t)-\hat w
C^{\alpha\sigma}_\beta(\zr_1+\r_1-\zr_2,t)]
C^{\gamma\gamma}_\sigma(\r_2,0)
$$
$$
=2\int{\d^3 k\over(2\pi)^3}{\d^3 p\over(2\pi)^3}{\d^3 q\over(2\pi)^3}
C_kC_pC_q\exp(-\eta_p|t|)W_\p(m_1)W_\p(m_2)
$$
$$
\times\Big\{{(\k\p)^2(\p\q)\over k^2p^2}-
{(\k\p)(\k\q)\over k^2}\Big\}
$$
$$
\times[(2-w_k(k+p))H(k+p-k_{n_1})-w_k(k+p)H(p-k_{n_1})]
$$
$$
\times[(2-w_p(p+q))H(p+q-k_{n_2})-w_p(k)H(q-k_{n_2})].\eqno({\rm B}2)
$$
The geometric term in braces is antisimmetric in $\q$, so that that only
nonzero contributions are proportional to $H(p+q-k_{n_2})$. We change first
variables: $q+p\to q$, so that $H(p+q-k_{n_2})\to H(q-k_{n_2})$. Because of the
factor $W_\p(m_1)W_\p(m_2)$, we can expand in a power series in $p$:
$$
-k_2^\alpha\partial_{k^\alpha_3}\Big[(kq)^{-5/3}
\Big({(\k\p)^2(\p\q)\over k^2p^2}-{(\k\p)(\k\q)\over k^2}\Big)\Big]=
-{5\over 3}(kq)^{-5/3}p^2(xyz-x^2z^2)
\eqno({\rm B}3)
$$
Indicating by $u$ the cosine of the angle between the planes $\k\p$ and $\k\q$,
we have:
$$
xyz-x^2z^2=uxz\sqrt{(1-x^2)(1-z^2)},$$
so that, substituting Eqn. (B3)
into (B2) gives zero. Going to next order produces a factor $p^2x^2$, which
gives zero again, while still the next order leads to a contribution which is
by a factor $(kR)^{2/3}$ smaller than the terms in the second group.

\vskip 5pt
The calculation of the terms in the secon group is very tedious but
straightforward and follows the same lines of Eqns. (28) and (B2). We conclude
by giving the exact form of the coefficients entering the expression for the
fluctuation amplitude $<g^2>$ $[$Eqn. (30)$]$:
$$
B_1(kpq)=-pk\,yz\,(y+xz);\qquad B_2(kpq)=kp\,(1+z^2)(z+xy);
$$
$$
B_3(kpq)=kp\,xy\,(1-z^2).\eqno({\rm B}4)
$$
$$
H_1=[(2-w_p(k))H(k-k_{n_1})-H(q-k_{n_1})w_p(k)]
$$
$$
\times[[(2-w_q(k))H(k-k_{n_2})-H(p-k_{n_2})w_q(k)]
$$
$$
-[(2-w_q(p))H(p-k_{n_2})-H(k-k_{n_2})w_q(p)]];\eqno({\rm B}5)
$$
$$
H_2=[(2-w_p(k))H(k-k_{n_1})-H(q-k_{n_1})w_p(k)]
$$
$$
\times[[(2-w_p(k))H(k-k_{n_2})-H(q-k_{n_2})w_p(k)]
$$
$$
-[(2-w_p(q))H(q-k_{n_2})-H(k-k_{n_2})w_p(q)]];\eqno({\rm B}6)
$$
$$
H_3=[(2-w_p(k))H(k-k_{n_1})-H(q-k_{n_1})w_p(k)]
$$
$$
\times[[(2-w_k(q))H(q-k_{n_2})-H(p-k_{n_2})w_k(q)]
$$
$$
-[(2-w_k(p))H(p-k_{n_2})-H(q-k_{n_2})w_k(p)]].\eqno({\rm B}7)
$$

\vfill\eject
\noindent{\bf Appendix C. Coefficients for the shell energy equation}
\vskip 5pt
\noindent The RHS of Eqn. (52) contains the following contributions:

\noindent A term giving the effect of the timescale becoming shorter at large
$n$:
$$
-F\int_0^\infty\d t\int\d n' g(n,n',t)\,F(n')\partial^2_{n'}\Xi(n',t)
$$
$$
=F\int_0^\infty\d t\int\d n' g(n,n',t)(1-t^2)<\xi_n(0)\xi_{n+n'}(t)>.
\eqno({\rm C}1)
$$
where:
$$
\Xi(n',t)={<\xi_n(0)\xi_{n+n'}(t)>\over<\xi_n(0)\xi_{n+n'}(0)>},\quad{\rm and}
\quad F(n')=<\xi_n(0)\xi_{n+n'}(0)>
\eqno({\rm C}2)
$$
A term coming purely from the noise, due to Beta Model effect:
$$
F\int_0^\infty\d t\int\d n' g(n,n'-n,t)<(\xi_n(n,0)-\xi_n(n-1,0))
(\xi_{n'}(n',t)-\xi_{n'}(n'-1,t))>
$$
$$
=F\beta\int_0^\infty\d t\int\d n' g(n,n',t)<\xi_n(0)\xi_{n+n'}(t)>;
\eqno({\rm C}3)
$$
where:
$$
\beta=1+a^{-3}-2^{5\over 2}(1+a^2)^{-{3\over 2}}.\eqno({\rm C}4)
$$
A cross correlation term between Beta Model effect and timescale mismatch:
$$
2F\int_0^\infty\d t\int\d n' g(n,n',t)\partial_n'\Xi(n',t)
<[\xi_n(n,0)-\xi_n(n-1,0)]\xi_{n+n'-1}(0)>
$$
$$
=F\beta'\int_0^\infty\d t\int\d n' g(n,n',t)<\xi_n(0)\xi_{n+n'-1}(t)>,
\eqno({\rm C}5)
$$
where:
$$
\beta'=2\gamma\Big(1-\Big({1+a^{-2})\over 2}\Big)^{3\over 2}\Big).
\eqno({\rm C}6)
$$
Finally there is the correction to the relaxation terms of the shell equation,
from Beta Model effect:
$$
(D-{V\over 2})<(\phi_n(n,0)-\phi_n(n-1,0))(\phi_n(n,t)-\phi_n(n-1,t))>
$$
$$
\simeq\beta(D-{V\over 2})<\phi_G^2>.\eqno({\rm C}7)
$$
To lowest order, there are no cross correlation between $(D-{V\over 2})
(\phi_G(n)-\phi_G(n+1))$ and previous terms due to the origin of $\phi_G$ from
$\partial_{n'}\xi$ rather than from $\xi$.

Using the formula $\int_0^\infty\d t\,\exp(-{b\over t^2}-ct^2)=\sqrt{\pi\over
4 c}\exp(-2\sqrt{bc})$ $[36]$, and shifting the $n'$ integration so that $n\to
0$, the integral term in Eqn. (52) can be rewritten in the form:
$$
\int\d n'\Big[<\xi_0(0)\xi_{n'}(0)>\Big[\gamma^2
\Big({\partial^2\over\partial c^2}-1\Big)
+\beta\Big]
$$
$$
+\beta'<\xi_0(0)\xi_{n'-1}(0)>{\partial\over\partial c}\Big]
\sqrt{\pi a'^2\over 4 c^2}\exp(d-2\sqrt{bc}),\eqno({\rm C}7)
$$
where:
$$
a'={1\over \sqrt{\pi D}};\qquad b={n'^2\over 4 D};
$$
$$
c={V\over 4D}+\exp(\gamma n');\qquad d={n'V\over 2 D}.
\eqno({\rm C}8)
$$
Writing explicitly:
$$
\int\d n'\Big\{\Big[\gamma^2\Big(1-{32D^2\over(V^2+4D)^2}-{6D|n'|\over(V^2+4D)^
{3\over 2}}-{n'^2\over V^2+4 D}\Big)+\beta\Big]{1\over 1+50(a-1)^2n'^2}
$$
$$
+\beta'{4D+|n'|\sqrt{4D+V^2}\over 4D+V^2}{1\over 1+50(a-1)^2(n'-1)^2}\Big\}
$$
$$
\times{F\over\sqrt{4D+V^2}}\exp\Big[{1\over 2D}(n'-|n'|\sqrt{4D+V^2})\Big].
\eqno({\rm C}9)
$$
It remains to evaluate the term $D<\phi\Delta^2\phi>$ on the LHS of Eqn. (51):
$$
<\phi\Delta^2\phi>=\int_0^\infty\d t_1\d t_2\int \d n_1\d n_2\,
F(n_1-n_2)g(n_1,t_1)\partial_{n_2}^2g(n_2,t_2)\,\Xi[t_1-t_2]
$$
$$
=\sum_{m=0}^\infty{\Xi_m\over m!}\int_0^\infty\d t\int\d n_1\d n_2\,
F(n_1-n_2)g(n_1,t_1)\partial_{t_2}^m\partial_{n_2}^2g(n_2,t_2),\eqno({\rm C}10)
$$
where
$$
\Xi_m\simeq\int\d t\,t^m\Xi(t).\eqno({\rm C}11)
$$
Using the defining relation for $g$: $(\partial_t-D\partial^2_n+V\partial_n)
g(n,t)=\delta(n)\delta(t)$ (in the continuum limit) and the condition:
$g(n,0)=g(n,\infty)=0$, we can write:
$$
<\phi\Delta^2\phi>=\sum_{m=0}^\infty{\Xi_m\over m!}\int_0^\infty\d t
\int\d n_1\d n_2\,F(n_1-n_2)g(n_1,t_1)(D\partial_{n_2}^2-V\partial_{n_2})^m
\partial_{n_2}^2g(n_2,t_2)
$$
$$
=\sum_{m=0}^\infty{\Xi_m\over m!}\int\d n_1\d n_2\,F(n_1-n_2)
(D\partial_{n_2}^2-V\partial_{n_2})^m
{n_1^2-n_2^2\over 2D(n_1^2+n_2^2)^2}.\eqno({\rm C}12)
$$
Integrating by part we obtain a series in the form:
$$
<\phi\Delta^2\phi>=\sum_{m=0}^\infty{\Xi_m\over m!}\int\d n_1\d n_2\,
{n_1^2-n_2^2\over 2D(n_1^2+n_2^2)^2}(a_m\partial_{n_2}^{2m}+b_m
\partial_{n_2}^{2m-1})F(n_1-n_2).\eqno({\rm C}13)
$$
Given the form of the correlation $F$ $[$see Eqn. (43)$]$, the action on it of
the operators $(a_m\partial_{n_2}^{2m}+b_m\partial_{n_2}^{2m-1})$ will lead to
terms in the form:
$$
{c_1+c_2(n_1-n_2)\over (1+50(a-1)^2(n_1-n_2)^2)^p}.\eqno({\rm C}14)
$$
Substituting back into Eqn. (C13), we see that terms in $c_1$ do not contribute
because of the antisimmetry of the integrand under the transformation
$n_1\to n_2$, $n_2\to n_1$. Similarly for the terms in $c_2$, due to the
antisimmetry under simultaneous change of sign of $n_1$ and $n_2$. Hence
the term $<\phi\Delta^2\phi>$ does not contribute to Eqn. (51).

\vfill\eject
\centerline{FIGURE CAPTIONS}
\vskip 5pt
\item{Fig. 1.} Kolmogorov constant $\Ckol$ and dimensionless parameter $\rho$,
as measured from the energy flux through scale $k$, in a reference frame moving
with velocity $\v(\lambda',k)=\hat w(\lambda',k)\v$, $0<\lambda'<0.9$, with
$\lambda=0.9$, fixed $[$ see Eqn. (14)$]$.
\vskip 5pt
\item{Fig. 2.} Energy transfer profiles $T(k,p)$ for $k=1$.
$(a)$: Lagrangean frame using Eqn. (15); $\lambda'=0.9$.
$(b)$: laboratory reference frame, using Eqn. (10).
$(c)$: Lagrangean frame using simplified closure with Eqn. (17) and:
$w_{p,q}=w_{p,q}(0.7,k)$. $(d)$: the same as $(c)$, but with $w_p=w_p(0.7,
\min(k,q))$ and $w_q=w_p(0.7,\min(k,p))$.
\vskip 5pt
\item{Fig. 3.} Normalized noise correlation ${<g_ng_{n+n'}>\over<g_n^2>}$
vs. shell separation $x=(a-1)n'$. $(a)$: calculation from Eqn. (30);
$(b)$: fit of $(a)$ by: $F(x)=(1+50x^2)^{-1}$.
$(c)$: profile for $A(x)=\int\d p |T(k,p)T(k+x,p+x)|$, normalized to $A(0)$;
\vskip 5pt
\item{Fig. 4.} Finite difference coefficients $A_r$ vs. $a$ $[$ see Eqns. (33),
(36) and (37)$]$. Notice that
higher order differences are always dominated by the advection-diffusion part.
\vskip 5pt
\item{Fig. 5.} Comparison of gaussian part of the fluctuation amplitude
$<<v_l^2>_\smalR^2>-<<v_l^2>^2>$, obtained from
integration of Eqn. (49) $(a)$, and from direct average of $v^\smalze$ $(b)$.
\vskip 5pt
\item{Fig. 6.} Scaling correction to $S(l,a,c_w,4)$, as a function of $a$ for
$c_w=2$; notice the saturation occurring at $a\simeq 1.3$.
\vskip 10pt
\item{Table B1.} Contractions contributing to $<g_n(m)g_{n'}(m')>$.
\vfill\eject
\end